%% file: main.tex
\documentclass[conference]{IEEEtran}
\usepackage{cite}
\usepackage{amsmath,amssymb,amsfonts}
\usepackage{graphicx}
\usepackage{tikz}
\usepackage{pgfplots}
\usepackage[mode=tex]{standalone} 
\usetikzlibrary{arrows.meta,positioning,shapes.geometric}
\usepgfplotslibrary{groupplots}
\usepgfplotslibrary{units}
\usepgfplotslibrary{fillbetween}
\pgfplotsset{compat=1.18}
\usepackage[group-four-digits=true,group-separator={,}]{siunitx}
\usepackage{textcomp}
\usepackage{hyperref}
\usepackage{cleveref}
\usepackage{subcaption}
\usepackage{capt-of}
\def\BibTeX{{\rm B\kern-.05em{\sc i\kern-.025em b}\kern-.08em
    T\kern-.1667eⁿm\lower.7ex\hbox{E}\kern-.125emX}}

\usepackage{etoolbox}
\newtoggle{releaseStuffAfterDoubleBlind}
\toggletrue{releaseStuffAfterDoubleBlind}
\newtoggle{releaseOnArxivWithoutAckno}
\togglefalse{releaseOnArxivWithoutAckno}
\newtoggle{makeReadyForSubmission}
\togglefalse{makeReadyForSubmission}
\newtoggle{inclIEEECopyRight}
\togglefalse{inclIEEECopyRight}
\newtoggle{includeAppendix}
\togglefalse{includeAppendix}
\iftoggle{releaseStuffAfterDoubleBlind}{%
    \toggletrue{makeReadyForSubmission}
}{%
    \usepackage[switch]{lineno} 
    
}
\iftoggle{makeReadyForSubmission}{%
    \usepackage{microtype}
}{%
    \usepackage{todonotes}
}
\iftoggle{makeReadyForSubmission}{%
    \newcommand{\revision}[1]{}
    \newcommand{\fixme}[1]{}
}{%
    \newcommand{\revision}[1]{{\color{green}{#1}}}
    \newcommand{\fixme}[1]{{\color{red}{#1}}}
}
\iftoggle{releaseOnArxivWithoutAckno}{%
    
}{%
    
}

\begin{document}

\ifboolexpr{ togl {releaseStuffAfterDoubleBlind} or togl {releaseOnArxivWithoutAckno} }{}{%
    \linenumbers
}

\title{Performance Portability of Out-of-Kernel Optimizations on GPUs: A Molecular Docking Case Study}

\noindent
{\small
\textbf{Note on this version.}
The performance results presented in this preprint are preliminary and were obtained using an earlier implementation of muDock.
An updated experimental evaluation is in preparation, and the reported measurements should not be considered representative of the current implementation.
}
\par\smallskip

\ifboolexpr{ togl {releaseStuffAfterDoubleBlind} or togl {releaseOnArxivWithoutAckno} }{%
    \author{
    	\IEEEauthorblockN{Gianmarco Accordi\IEEEauthorrefmark{1}, Leonardo Beltrame\IEEEauthorrefmark{1},
    		Theresa Pollinger\IEEEauthorrefmark{2},
    		Jens Domke\IEEEauthorrefmark{2},
    		Davide Gadioli\IEEEauthorrefmark{1},
    		Gianluca Palermo\IEEEauthorrefmark{1}}
    
    	\IEEEauthorblockA{\IEEEauthorrefmark{1}Dipartimento di Elettronica, Informazione e Bioingegneria, Politecnico di Milano, Milan, Italy\\
    		\{gianmarco.accordi, leonardo.beltrame, davide.gadioli, gianluca.palermo\}@polimi.it}
    
     	\IEEEauthorblockA{\IEEEauthorrefmark{2}RIKEN Center for Computational Science, Kobe, Japan\\
     		\{jens.domke, theresa.pollinger\}@riken.jp}
    }%
}{\author{}}%

\maketitle

\begin{abstract}
Achieving performance portability is challenging for irregular scientific applications built as repeated multi-kernel GPU pipelines, because throughput depends on out-of-kernel decisions as much as on kernel efficiency. 
Using molecular docking as a representative case study, we propose an out-of-kernel tuning methodology that treats batching policy, worker concurrency, and per-worker memory allocation as explicit optimization parameters, replacing backend-dependent occupancy-driven choices with a more portable execution strategy based on explicit resource bounds. 
We implement this methodology in the muDock mini-app under a manager/worker execution model where multiple host workers share a GPU through independent streams. We then evaluate the methodology on NVIDIA, AMD, and Intel GPUs across CUDA, HIP, and SYCL implementations. 
Across all evaluated platform/backend pairs, the memory-bounded formulation exceeds the peak throughput of the occupancy-guided baseline. The fully tuned methodology improves the end-to-end throughput on all tested configurations, achieving an average speedup of 1.53$\times$ over the baseline.

\end{abstract}


\input{sections/introduction}
\input{sections/background}
\input{sections/methodology}
\input{sections/experimental_setup}
\input{sections/experimental_result}
\input{sections/discussion}
\input{sections/related_work}
\input{sections/conclusion}

\section*{Acknowledgment}
The authors take full responsibility and ownership for the paper content, while ChatGPT and DeepL have been used for language editing.

\section*{Data Availability}
The source code, benchmarking scripts, and datasets used in this study will be made publicly available upon acceptance via GitHub and Zenodo.

\bibliographystyle{IEEEtran}
\bibliography{references}

\iftoggle{includeAppendix}{
\appendices
\input{sections/appendix_plots}
}{}

\end{document}

%% file: sections/introduction.tex
\section{Introduction}
\label{sec:introduction}

High-performance computing (HPC) plays a central role in reducing the time-to-solution of large-scale scientific applications.
As production systems increasingly rely on GPU acceleration, however, achieving this objective across different vendors and software ecosystems becomes more difficult in practice.
Deploying a scientific application across platforms requires not only functional portability, but also tuning choices that remain effective despite differences in compilers, runtimes, and hardware resource behavior~\cite{pennycook2019implications,pennycook2021revisiting,crisci2024ligen_pipeline_pp}.
This problem is particularly relevant for applications that are not dominated by a single monolithic kernel, but instead execute long sequences of repeated kernel launches.
In these applications, throughput depends on how inputs are grouped and submitted across the full pipeline, not only on the efficiency of one kernel in isolation.
As a result, achieving portable performance requires looking at the full application pipeline, not only at kernel-level code generation.

As an example, biological applications consume a significant fraction of HPC resources, and virtual screening is a representative throughput-oriented workload in which a speed-up directly increases the screening capacity.
Structure-based virtual screening uses docking to rank large libraries of compounds, called \textit{ligand}s, against one or more protein targets, allowing researchers to focus downstream experimental validation on the most promising candidates~\cite{shoichet2004virtual,kitchen2004docking,lionta2014sbvs}.
Because docking simulations are much cheaper than experimental screening, modern campaigns increasingly target very large chemical libraries, sometimes reaching hundreds of millions of compounds~\cite{gadioli2021exscalate,lyu2019ultralarge,liu2025impact}.
This trend makes portable GPU efficiency a first-order concern for docking pipelines.
Docking is also a challenging workload for portable performance.
Ligands differ in atom count, flexibility, and interaction patterns, so the cost and memory footprint of processing each ligand can differ substantially~\cite{vitali2024gpu_optimized_docking}.
This irregularity makes batching, resource usage, and concurrent device use especially important for end-to-end throughput.
Prior work on out-of-kernel optimizations has already shown that these decisions can be as important as kernel-level tuning~\cite{accordi2024ook}.

In this paper, we study this problem using \texttt{muDock}~\cite{accordi2025mudock_cpu,mudock_repo}, a compact molecular docking mini-app designed to preserve the key computational patterns of larger docking codes while enabling systematic experimentation across platforms and backends.
Our goal is to study whether a more portable out-of-kernel formulation can improve throughput in a repeated multi-kernel GPU pipeline.
Instead of relying on occupancy-driven batch sizing, which depends on backend-specific information and assumptions, we study an alternative formulation based on explicit resource bounds.
In particular, we focus on batching policy, worker concurrency, and per-worker memory allocation as coupled tuning dimensions, and evaluate how these choices transfer across CUDA, HIP, and SYCL implementations.

This work makes the following contributions:
\begin{itemize}
	\item We study an out-of-kernel tuning approach for repeated multi-kernel GPU pipelines based on explicit resource bounds, treating batching policy, worker concurrency, and per-worker memory allocation as the main tuning dimensions.
	\item We implement and evaluate this approach in \texttt{muDock} on NVIDIA GH200, AMD MI250, and Intel Max 1550 systems using CUDA, HIP, and SYCL, showing that the tuned configuration improves end-to-end throughput by 1.53$\times$ on average over the reference configuration.
\end{itemize}

The rest of the paper is organized as follows.
Section~\ref{sec:background} introduces the workload and portability context.
Section~\ref{sec:methodology} presents the batching policies and tuning dimensions studied in this work.
Sections~\ref{sec:experimental_setup} and~\ref{sec:experimental_result} describe the experimental setup and the main results.
Section~\ref{sec:discussion} discusses limitations and threats to validity.
Section~\ref{sec:related_work} summarizes the most relevant prior work.
Section~\ref{sec:conclusion} concludes the paper.

%% file: sections/background.tex
\begin{figure*}[th!]
\centering
\resizebox{\textwidth}{!}{
\begin{tikzpicture}[
  node distance=8mm and 9mm,
  box/.style={draw, rounded corners, align=center, inner sep=4pt},
  io/.style={draw, circle, align=center, inner sep=4pt},
  decision/.style={draw, diamond, aspect=2, align=center, inner sep=2pt},
  arr/.style={-{Latex[length=2mm]}, thick}
]
\node[io] (in) {Input ligands};
\node[box, right=of in] (g0) {GENETIC kernel\\initialization/update};
\node[box, right=of g0] (geom) {GEOM kernel\\pose/transform stage};
\node[box, right=of geom] (score) {SCORE kernel\\fitness evaluation};
\node[box, right=of score] (g1) {GENETIC kernel\\selection/mutation};
\node[decision, right=of g1] (stop) {Stop\\criterion?};
\node[io, right=of stop] (out) {Output};

\draw[arr] (in) -- (g0);
\draw[arr] (g0) -- (geom);
\draw[arr] (geom) -- (score);
\draw[arr] (score) -- (g1);
\draw[arr] (g1) -- (stop);
\draw[arr] (stop) -- node[above]{yes} (out);
\draw[arr] (stop.south) |- ++(0,-11mm) -| node[pos=0.25,below]{no (repeat)} (g0.south);
\end{tikzpicture}
}
\caption{Overview of the genetic algorithm that \texttt{muDock} uses to dock a bucket of ligands. The rounded boxes represent kernel invocations, the circles represent the input and output of the algorithm, and the rhombus represents the termination condition.}
\label{fig:pipeline_loop_overview}
\end{figure*}

\section{Background}
\label{sec:background}


Drug discovery is a long and costly process with a high failure rate.
One of the challenges is the sheer size of the chemical space that we can explore to find a promising drug candidate \cite{polishchuk2013estimation}.
To mitigate this problem, a recent trend introduces an in-silico step that virtual screen a chemical library of ligands, to find the most promising ones to test in-vitro.
Recent studies demonstrate how increasing the number of virtual screened ligands, increases the probability of finding good drug candidates as well \cite{liu2025impact}.
In the context of a virtual screening campaign, it is of paramount importance to maximize the number of virtual screened ligands within the computation budget.
For this reason, we can capitalize on every optimization opportunity by evaluating more ligands, instead of reducing the time to solution, making virtual screening a relevant example of an irregular throughput-oriented workload.

To study this problem in a controlled setting, we use \texttt{muDock} as a compact molecular docking mini-app designed for benchmarking and portability studies.
It preserves the main computational structure of larger docking codes while remaining compact enough to support systematic experimentation across platforms and backends.

From the application point of view, \texttt{muDock} docks each ligand independently from the others, making the workload embarrassingly parallel.
Docking a ligand means estimating the 3D displacement of its atoms when they interact with the target protein.
To reach its goal it uses a genetic algorithm approach.
Each individual's genes encode geometric transformations to the ligands atoms.
In particular, from the initial ligand pose, we encode in the first six genes the rigid rotations and translations along the x, y, and z axis.
The other genes encode the angles of the ligand's rotatable bonds.
To evaluate the fitness of a population individual, \texttt{muDock} uses a scoring function that is functionally equivalent to the AutoDock force field described by Huey et al.~\cite{huey2007autodock_ff}.
As the population evolves, we expect the best individuals to converge toward the scoring function maximum.

\Cref{fig:pipeline_loop_overview} shows how the genetic algorithm maps with computation kernels.
The idea is to compose a computation pipeline to produce the best 3D atom displacement in output, starting from the initial pose.
The genetic stage initializes and updates the candidate poses explored during docking.
The geometry stage applies the corresponding pose transformations to the ligand coordinates.
The scoring stage evaluates the transformed poses and returns the fitness values used by the next genetic iteration.
After each iteration, the pipeline returns to the genetic stage until the stopping criterion is met.

\texttt{muDock} uses a manager/worker execution model.
In this context, a worker is a host process pinned to a CPU core and associated with one execution device selected by the user from the command line.
The device can be either a GPU or a CPU core.
For a GPU, the runtime can spawn multiple workers for the same device.
For a CPU execution path, the runtime instead uses one worker for each CPU core.
In this paper, it is assumed that the main compute power is provided by the GPUs of the node, and assigning workers on CPU is omitted.

We use this model as a straightforward way of tightly coupling the different resources (data, compute, memory).
Of course, there is a multitude of other approaches in the context of task-based runtime systems and performance portability frameworks to coordinate different kinds of resources, and even combinations~\cite{daissForkJoinIntegrationPerformance2021}.
However, the molecular docking workload allows for a straightforward one-dimensional partitioning of all the work, by grouping into batches. 
There are no data dependencies that warrant a complex task graph.
Thus, by focusing on the manager/worker model, one can avoid unnecessary synchronization and data movement: every worker is responsible for a single GPU stream.
Still, the presented insights should be largely transferable to the wider context of task-based runtime systems on GPU workloads.

%% file: sections/methodology.tex
\section{Methodology}
\label{sec:methodology}

This section describes the out-of-kernel knobs that we investigate in this paper.
While they are evaluated in the context of \texttt{muDock}, they are agnostic with respect to the mechanism used to orchestrate the computation.
At a high level, our study focuses on two complementary aspects of the out-of-kernel execution strategy.
The first aspect concerns how the work is grouped into batches under explicit resource constraints.
The second aspect concerns how the workers coordinate and submit these batches to the GPU to overlap data transfer with computation and ensure that the pipeline remains busy.
In this work, these two aspects are jointly analyzed rather than in isolation, because the effect of batch sizing depends on how much work can be kept in flight through concurrent submission and memory partitioning across workers.

\subsection{Batch sizing}\label{sec:method_batch_sizing}

Grouping a set of inputs in a batch is an efficient way of using a GPU when we have data parallelism~\cite{vitali2024gpu_optimized_docking}.
In this work, we investigate two different methods to automatically size the batch.

\subsubsection{Occupancy-Guided Batching}
\label{sec:method_batch_construction}

The first method considered in this paper is occupancy-guided batching.
Its goal is to form batches that keep the GPU fully occupied while grouping together ligands with similar characteristics, so that the ligands inside the same batch have a similar expected computation time.
This follows the same rationale used in earlier docking work based on occupancy-guided execution~\cite{accordi2024ook}.

Ligands are first grouped into classes according to their input features.
For each class $c$, the runtime estimates the per-ligand memory footprint $m_{\text{lig}}(A_c)$.
Given the memory budget assigned to a worker, the runtime determines the largest safe batch size,
\begin{equation}
B_{\max}(c)=\left\lfloor\frac{M_{\text{avail}}}{m_{\text{lig}}(A_c)}\right\rfloor,
\end{equation}
where $M_{\text{avail}}$ is the memory budget assigned to the worker.
This value is an estimated upper bound imposed by correctness and memory availability.

In addition to $B_{\max}$, the runtime derives an occupancy-related base multiple from backend helpers~\cite{accordi2024ook}.
For one worker, let $a_{\text{sm}}$ be the estimated number of active blocks per SM or compute unit returned from the backend, and let $n_{\text{sm}}$ be the number of SMs or compute units on the device.
The value of $a_{\text{sm}}$ is kernel dependent, since it reflects how the resource usage of the selected kernel maps onto the hardware limits of the target device.
The occupancy-derived base multiple is then
\begin{equation}
B_{\text{base}}=\max(1,a_{\text{sm}})\cdot\max(1,n_{\text{sm}}).
\end{equation}
The runtime can then use integer multiples of this quantity to enlarge the batch while preserving the same occupancy-related alignment, as long as the resulting batch still fits within the available memory budget.

In a repeated multi-kernel pipeline, however, the occupancy-related base multiple is not unique.
Different kernels may expose different preferred multiples.
In the experiments, we combine these hints conservatively by selecting the minimum value across the relevant kernels.
This means that the batch is dimensioned for the worst case, i.e., for the kernel with the highest resource usage.
This is appropriate for the limiting kernel, but it can lower the occupancy reached by the other kernels in the pipeline.
Given this limitation, we also consider a simpler method which is agnostic with respect to the occupancy, as it considers only the memory occupancy of batches, that we describe in the following section.


\subsubsection{Memory-Bounded Batching}
\label{sec:method_bucket_resolution}

The second method is memory-bounded batching.
Here, the runtime still groups ligands by class and still respects the same memory-safety condition, but it resolves the batch size directly from the worker memory budget,
\begin{equation}
B=B_{\max}.
\end{equation}
In this case, the full memory budget assigned to a worker is used to form the batch, without requiring alignment to an occupancy-derived multiple.

The purpose of this method is different from the occupancy-guided one.
Instead of sizing the batch for the most constraining kernel in the pipeline, it tries to increase the amount of work in flight by allowing larger batches and concurrent workers to use the device together.
This is particularly relevant for irregular multi-kernel workloads, where the different stages do not necessarily share the same preferred batch multiple.

Under this method, the number of workers $W$ and the memory budget $M$ assigned to each worker become explicit tuning parameters.
Each worker uses its own memory budget to build the next batch and submits work through an independent stream.
The method therefore relies on worker concurrency and larger in-flight batches to improve overlap and to help hide latencies across the pipeline.

\subsection{Kernel execution policy}

Computation efficiency on GPU is usually achieved by overlapping data transfer with computation.
It is common to use $W$ host workers that enqueue kernels on the same device.
In \texttt{muDock}, a manager owns the shared input backlog and the shared output queue.
The input backlog contains the ligand descriptions that still need to be processed, while the output queue collects the completed results produced by the workers.
The manager spawns one or more workers according to the selected devices and backends.
Each worker repeatedly removes work descriptions from the shared input backlog, prepares the corresponding input data, and stores it in a local reordering buffer.
The role of this local buffer is to group inputs with similar features before forming the next batch, reducing the imbalance inside the same device submission.
When the inputs accumulated in the local buffer satisfy the batching rule of the selected method, the worker forms a batch and submits it to the device.
The worker then executes the target pipeline for all inputs in that batch and writes the completed results to the shared output queue.
This process continues until the input backlog has been drained.

This execution model exposes two levels of concurrency.
Inside the device, each submitted batch is processed by a sequence of GPU kernels.
Outside the kernels, multiple host workers may concurrently prepare inputs, reorder them, form batches, and submit work to the same device.
As a consequence, end-to-end throughput depends not only on the behavior of the computational stages, but also on how work is distributed across workers, how inputs are grouped in the local buffers, and when each worker decides that the next batch is ready for execution.
However, the two methods that we investigate in this paper to overlap data transfer with computation are agnostic with respect to the worker implementation.
This setup exposes the main tradeoff studied in the paper.
Increasing $W$ can improve overlap and reduce launch-related idle time, but it also reduces the memory $M$ available to each worker and may increase contention.
For this reason, the worker count and the per-worker memory budget are part of the methodology itself rather than fixed background settings.

\subsubsection{Explicit synchronization}

An established approach uses explicit synchronization between two workers to implement double buffering~\cite{vitali2024gpu_optimized_docking}.
The idea is that while one thread is executing a computation kernel on a buffer, another worker is transferring data to a second buffer.
Once the kernel is completed, the second worker can issue its own computation kernel and the first worker can transfer data.
This can be easily generalized to $W$ workers.
The benefit is that with this approach the developer is certain to overlap data transfer and computation regardless of the backend used to offload computation.
The drawback is that a worker needs to wait for another kernel to complete before launching its own kernel.
For this reason, this policy is usually paired with occupancy-guided batching~\cite{accordi2024ook} to maximize the computation efficiency.

\subsubsection{Concurrent execution}

An alternative way relies on GPU streams and the underlying driver to overlap data transfer with computation.
With this approach, each worker enqueues data transfer and computation kernels in a stream independently.
For example, in case of the CUDA backend, a stream is represented by the CUDA stream.
However, also the other backends expose to the developer a similar runtime to launch kernel concurrently.
The benefit is that the resulting implementation is simpler, and it is possible to overlap data transfer and computation across the whole execution.
The drawback is that the ability of maximizing the efficiency is up to the backend runtime and the underlying GPU driver.
This policy can be paired with memory-bounded batching to maximize the number of in-flight ligands on the device.





%% file: sections/experimental_setup.tex
\section{Experimental Setup}
\label{sec:experimental_setup}

This section provides the details about the platforms that we used to carry out the experimental campaign, the input dataset, and how we measure the throughput.

\paragraph{Hardware platforms}
The campaign spans three production GPU platforms from different vendors.
The NVIDIA GH200 measurements are collected on a node with a 72-core Arm Neoverse-V2 host processor, \qty{477}{\giga\byte} of shared system memory, and one NVIDIA GH200 GPU.
The AMD MI250 measurements are collected on a node with two AMD EPYC 7713 processors, \qty{1024}{\giga\byte} of system memory, and eight AMD MI250 GPUs with \qty{128}{\giga\byte} of memory each.
The Max1550 measurements are collected on a node with two Intel Xeon Platinum 8470 processors, 208 logical CPU cores in total, \qty{2}{\tera\byte} of system memory, and eight Intel Data Center GPU Max 1550 accelerators with \qty{128}{\giga\byte} of memory each.

\paragraph{Software stack}
All nodes use Linux-based software environments.
The GH200 and Max1550 systems run Rocky Linux~9.6 with Linux~5.14 kernels, and GCC~13.2.0 is used as the host compiler across the campaign.
On GH200, the native CUDA implementation is built with CUDA~12.6.2 and NVIDIA driver~590.48.01, while the HIP-on-NVIDIA configuration also resolves to the same CUDA~12.6.2 toolchain underneath.
The SYCL builds use an Intel LLVM/oneAPI-based compiler stack: on GH200 they target NVIDIA GPUs, on MI250 they target AMD \texttt{gfx90a}, and on Max1550 they target \texttt{intel:pvc}.
On Max1550, the SYCL runtime uses Intel oneAPI Unified Runtime over Level Zero~1.6.33578+42, with OpenCL NEO~25.18.33578 also available on the system.
For AMD HIP, we use the ROCm~6.2.1 LLVM/AMDGPU toolchain targeting \texttt{gfx90a}, together with the corresponding \texttt{rocrand} and \texttt{hiprand} libraries.
The Intel LLVM SYCL compiler used in the study was built from source from the main branch at commit \texttt{a123482f3}. 

\paragraph{Workloads}
The application performance is highly input-dependent, as the execution time is strongly related to the number of heavy atoms and rotamers \cite{gadioli2021exscalate}.
For this reason, the study focuses on the heterogeneous \texttt{mixed\_n500k} workload, built as a subsample of \num{500000} compounds from the MEDIATE dataset \cite{Vistoli03082023}.
The selected ligands are representative of different combinations of heavy-atom counts and rotamer counts, so that the workload preserves the irregular behavior that drives the batching problem.
At the same time, the sampled distribution is intentionally biased toward a higher presence of smaller ligands than larger ones, to better reflect a common screening scenario in which lighter compounds are more frequent than the most complex molecules.
All policy sweeps, headline comparisons, and portability results discussed in the paper are therefore derived from this mixed workload.

\paragraph{Throughput metric}
Throughput is the main performance metric of interest for the end user, since it directly estimates how many ligands can be screened within a fixed time budget.
Runs are monitored over time through observer samples.
For each experiment, we report the throughput measured when the input backlog has been drained, namely when all ligands have either already completed or are in flight inside the pipeline.
This choice lets us compare experiments that complete naturally with experiments that are stopped after a fixed timeout, without penalizing runs whose input queue empties early.
We introduce a timeout of 5 minutes to keep the campaign duration reasonable, after observing that the pipeline reaches a stable saturation regime well before that limit in the valid configurations.
Each experiment is executed five times; we discard the first run as warm-up and average the remaining four.
If one of the five repetitions fails, for example, because the selected configuration exceeds the available memory budget, we discard the entire set of repetitions and mark that experiment as not available (N/A).
As a consequence, missing points in the worker/memory and bucket-multiplier plots correspond to configurations for which at least one repetition could not be completed, typically because the requested batching and worker configuration required more device memory than the device could provide.

\paragraph{Performance Portability Metrics}
\label{sec:method_pp_metrics}

To summarize the tuned results across platforms and backends, we use the application-efficiency view adopted from the \texttt{muDock} study on CPU performance portability~\cite{accordi2025mudock_cpu} and grounded in the discussion of Pennycook and Sewall~\cite{pennycook2019implications,pennycook2021revisiting}.
For a platform $p$ and a supported backend $b$, the application efficiency is
\begin{equation}
e(b,p)=\frac{T(b,p)}{\max_{b' \in B_p} T(b',p)},
\end{equation}
where $T(b,p)$ is the measured throughput and $B_p$ is the set of backends supported on platform $p$.
This metric reports how close a backend is to the best tuned result available on the same platform.

We then summarize these efficiencies across platforms using the arithmetic mean and the harmonic mean.
For a backend supported on $N$ platforms, the arithmetic mean is
\begin{equation}
\mathrm{AM}(b)=\frac{1}{N}\sum_{p} e(b,p),
\end{equation}
while the harmonic mean is
\begin{equation}
\mathrm{HM}(b)=\frac{N}{\sum_{p} 1/e(b,p)}.
\end{equation}
The arithmetic mean gives a simple average-efficiency view.
The harmonic mean is stricter because it penalizes low-efficiency cases more strongly and becomes zero when a backend is unsupported on one of the investigated platforms.
For this reason, we use the harmonic mean as the main portability-oriented summary, while reporting the arithmetic mean as a complementary metric.

%% file: sections/experimental_result.tex
\begin{figure*}[t]
\centering
\begin{tikzpicture}[baseline]
\matrix[column sep=4mm, row sep=0mm] {
\draw[blue!80!black, line width=1.1pt] (0,0) -- (0.35,0); \fill[blue!80!black] (0.175,0) circle (1.4pt); & \node[font=\footnotesize] {$W=1$}; &
\draw[teal!80!black, line width=1.1pt] (0,0) -- (0.35,0); \filldraw[teal!80!black] (0.175,0) rectangle +(0.01,0.01); & \node[font=\footnotesize] {$W=2$}; &
\draw[green!50!black, line width=1.1pt] (0,0) -- (0.35,0); \fill[green!50!black] (0.175,0) -- +(0.018,0.031) -- +(-0.018,0.031) -- cycle; & \node[font=\footnotesize] {$W=4$}; &
\draw[orange!90!black, line width=1.1pt] (0,0) -- (0.35,0); \fill[orange!90!black] (0.175,0) -- +(0.02,0.02) -- +(0.04,0) -- +(0.02,-0.02) -- cycle; & \node[font=\footnotesize] {$W=8$}; &
\draw[magenta!80!black, line width=1.1pt] (0,0) -- (0.35,0); \fill[magenta!80!black] (0.175,0) circle (1.6pt); & \node[font=\footnotesize] {$W=16$}; \\
};
\end{tikzpicture}
\vspace{1mm}

\begin{subfigure}{0.49\textwidth}
    \centering
    \includestandalone[width=\linewidth]{pictures/tikz/bucket_mult_gh200_cuda}
    \caption{GH200/CUDA}
    \label{fig:gh200-cuda}
\end{subfigure}
\hfill
\begin{subfigure}{0.49\textwidth}
    \centering
    \includestandalone[width=\linewidth]{pictures/tikz/bucket_mult_gh200_hip}
    \caption{GH200/HIP}
    \label{fig:gh200-hip}
\end{subfigure}
\hfill
\begin{subfigure}{0.49\textwidth}
    \centering
    \includestandalone[width=\linewidth]{pictures/tikz/bucket_mult_gh200_sycl}
    \caption{GH200/SYCL}
    \label{fig:gh200-sycl}
\end{subfigure}
\hfill
\begin{subfigure}{0.49\textwidth}
    \centering
    \includestandalone[width=\linewidth]{pictures/tikz/bucket_mult_mi250_hip}
    \caption{MI250/HIP}
    \label{fig:mi250-hip}
\end{subfigure}
\hfill
\begin{subfigure}{0.49\textwidth}
    \centering
    \includestandalone[width=\linewidth]{pictures/tikz/bucket_mult_mi250_sycl}
    \caption{MI250/SYCL}
    \label{fig:mi250-sycl}
\end{subfigure}
\hfill
\caption{Application throughput by varying the Bucket Multiplier on the x-axis, leading to larger batches toward right values. The curves correspond to worker counts, and the two panels compare the occupancy-guided method under two different GPU access methods.}
\label{fig:bucket-mult-gh200-cuda}
\end{figure*}
\section{Experimental Results}
\label{sec:experimental_result}

This section reports four experiments.
We first study the occupancy-guided approach by varying the Bucket Multiplier, and therefore the batch size, together with the number of active workers.
We then compare occupancy-guided and memory-bounded batching under different worker counts and per-worker memory budgets.
Next, we report the end-to-end throughput gain of the Tuned configuration over the Baseline configuration.
Finally, we summarize the tuned results through application-efficiency and performance-portability metrics.

\subsection{Bucket Multiplier and GPU Access Mode}
\label{sec:exp_bucket_mult}

This experiment studies the occupancy-guided approach.
The goal is to evaluate whether this approach still holds in a repeated multi-kernel pipeline.
In particular, we study whether explicit synchronization, implemented through kernel locking, remains beneficial across different architectures and backends.
For each backend-architecture pair, we measured the application throughput by varying the memory budget and number of workers.
Figure~\ref{fig:bucket-mult-gh200-cuda} reports throughput in ligands/\unit{\second} as a function of the Bucket Multiplier.
The curves correspond to different worker counts, and, for each platform/backend pair, the two panels compare the two GPU access modes.
\emph{Locked GPU Access} means that one worker uses explicit synchronization, whereas \emph{Concurrent GPU Access} means that each worker relies on its own GPU stream.
We do not include Max1550/SYCL in this figure because the occupancy-guided method depends on backend information about the number of active units per kernel, and on Intel GPUs this quantity was returned through SYCL as a default value equivalent to using the whole compute unit regardless of the kernel under analysis.

If the monolithic occupancy-guided assumption remained valid across multiple kernels, the two methods would stay close even as the number of workers increases.
This is not the case.
With one worker, the two configurations remain close, but they diverge once multiple workers are active.
When locked GPU access is enabled, the workers cannot keep enough work in flight and the attainable occupancy remains lower.
By contrast, under concurrent GPU access, the occupancy-guided approach already benefits from concurrent submissions and reaches a broader high-throughput region.
The same trend also appears along the Bucket Multiplier axis: once the GPU is no longer used in exclusive mode, fewer workers are sufficient to sustain high throughput even at lower Bucket Multiplier values.
In practice, concurrent GPU access reduces the need to rely on a large occupancy-aligned batch to keep the device busy, because overlap across workers already increases the amount of work in flight.

\subsection{Batching Policies Under Worker and Memory Scaling}
\label{sec:exp_memory_only_vs_aligned}

This experiment compares occupancy-guided batching and memory-bounded batching under explicit worker and memory partitioning.
The goal is to measure how the two batching methods behave on an irregular repeated multi-kernel workload.
Figure~\ref{fig:mem-work-all} reports throughput in ligands/\unit{\second} on the y-axis and the memory budget assigned to each worker on the x-axis.
The curves correspond to different worker counts, and each subplot reports one platform/backend pair.
Within each subplot, the left panel reports the occupancy-guided version and the right panel reports the memory-bounded version; the same worker counts are shown in both panels to make the comparison direct.

\begin{figure*}[t]
\centering
\begin{tikzpicture}[baseline]
\matrix[column sep=4mm, row sep=0mm] {
\draw[blue!80!black, line width=1.1pt] (0,0) -- (0.35,0); \fill[blue!80!black] (0.175,0) circle (1.4pt); & \node[font=\footnotesize] {$W=1$}; &
\draw[teal!80!black, line width=1.1pt] (0,0) -- (0.35,0); \filldraw[teal!80!black] (0.175,0) rectangle +(0.01,0.01); & \node[font=\footnotesize] {$W=2$}; &
\draw[green!50!black, line width=1.1pt] (0,0) -- (0.35,0); \fill[green!50!black] (0.175,0) -- +(0.018,0.031) -- +(-0.018,0.031) -- cycle; & \node[font=\footnotesize] {$W=4$}; &
\draw[orange!90!black, line width=1.1pt] (0,0) -- (0.35,0); \fill[orange!90!black] (0.175,0) -- +(0.02,0.02) -- +(0.04,0) -- +(0.02,-0.02) -- cycle; & \node[font=\footnotesize] {$W=8$}; &
\draw[magenta!80!black, line width=1.1pt] (0,0) -- (0.35,0); \fill[magenta!80!black] (0.175,0) circle (1.6pt); & \node[font=\footnotesize] {$W=16$}; \\
};
\end{tikzpicture}
\vspace{1mm}

\begin{subfigure}{0.49\textwidth}
    \centering
    \includestandalone[width=\linewidth]{pictures/tikz/mem_work_gh200_cuda}
    \caption{GH200/CUDA}
    \label{fig:gh200-cuda}
\end{subfigure}
\hfill
\begin{subfigure}{0.49\textwidth}
    \centering
    \includestandalone[width=\linewidth]{pictures/tikz/mem_work_gh200_hip}
    \caption{GH200/HIP}
    \label{fig:gh200-hip}
\end{subfigure}
\hfill
\begin{subfigure}{0.49\textwidth}
    \centering
    \includestandalone[width=\linewidth]{pictures/tikz/mem_work_gh200_sycl}
    \caption{GH200/SYCL}
    \label{fig:gh200-sycl}
\end{subfigure}
\hfill
\begin{subfigure}{0.49\textwidth}
    \centering
    \includestandalone[width=\linewidth]{pictures/tikz/mem_work_mi250_hip}
    \caption{MI250/HIP}
    \label{fig:mi250-hip}
\end{subfigure}
\hfill
\begin{subfigure}{0.49\textwidth}
    \centering
    \includestandalone[width=\linewidth]{pictures/tikz/mem_work_mi250_sycl}
    \caption{MI250/SYCL}
    \label{fig:mi250-sycl}
\end{subfigure}
\hfill
\begin{subfigure}{0.49\textwidth}
    \centering
    \includestandalone[width=\linewidth]{pictures/tikz/mem_work_pvc_sycl}
    \caption{Max1550/SYCL}
    \label{fig:pvc-sycl}
\end{subfigure}

\caption{Application throughput by varying the memory budget assigned to each worker (x-axis), the application throughput in ligands per second, and the number of host workers (curves). Each subplot reports one platform/backend pair; within each subplot, the left panel reports the occupancy-guided version and the right panel reports the memory-bounded version.}
\label{fig:mem-work-all}
\end{figure*}

The results show that memory-bounded batching is better suited to this multi-kernel execution model, it provides equal or better peak throughput and a broader useful operating region.
This is the main trend visible in the plots.
Across the evaluated platform/backend pairs, memory-bounded batching matches or exceeds the peak throughput of occupancy-guided batching and avoids dependence on backend-specific occupancy estimates.

\subsection{End-to-End Throughput Improvement}
\label{sec:exp_headline_baseline_tuned}

\begin{figure*}[t]
\centering
\includestandalone[width=0.97\textwidth]{pictures/tikz/speedups_mixed}
\caption{Cross-platform mixed-workload throughput. The x-axis reports the tested backends grouped by platform, the y-axis reports throughput in ligands/\unit{\second}, the bars compare the reference configuration and the tuned methodology, and the labels above the tuned bars report speedup over the reference.}
\label{fig:headline_baseline_tuned}
\end{figure*}

This experiment compares the \emph{Tuned} and \emph{Baseline} configurations on the mixed workload as summarized in Table~\ref{tab:tuned-runtime-configs}.
The goal is to measure the end-to-end effect of tuning batching, worker count, memory allocation, and runtime configuration together.
In this context, the \emph{Baseline} is the approach proposed in literature~\cite{accordi2024ook}, which used occupancy-guided batching with explicit synchronization.
It uses double buffering, a Bucket Multiplier equal to 3, and two workers on all supported platform/backend pairs.
\emph{Tuned} denotes the memory-bounded configuration with concurrent GPU access, using the worker count and per-worker memory budget selected from the worker/memory study for each platform/backend pair.
We report the result of the experiment in Figure~\ref{fig:headline_baseline_tuned}.
\emph{Baseline} denotes the fixed occupancy-guided reference configuration used in the comparison.

\begin{table}[t]
\centering
\small
\caption{Tuned runtime configurations selected from the worker/memory experiment and used for the Tuned bars in Figure~\ref{fig:headline_baseline_tuned}.}
\label{tab:tuned-runtime-configs}
\begin{tabular}{lcc}
\hline
Platform/Backend & Workers & Memory [GB] \\
\hline
GH200/CUDA & 4 & 8 \\
GH200/HIP & 8 & 4 \\
GH200/SYCL & 4 & 16 \\
MI250/HIP & 8 & 4 \\
MI250/SYCL & 4 & 8 \\
Max1550/SYCL & 8 & 8 \\
\hline
\end{tabular}
\end{table}

The results show that the \emph{Tuned} configuration improve throughput across the tested vendor/framework combinations.
This is what Figure~\ref{fig:headline_baseline_tuned} shows.
The \emph{Tuned} configuration improves the mixed-workload throughput on all evaluated platform/backend pairs, with a minimum/mean/maximum speedup of 1.38$\times$/1.53$\times$/1.83$\times$ over Baseline.
The comparison between \emph{Baseline} and \emph{Tuned} is intentionally practical rather than fully controlled: the former is a fixed occupancy-guided reference point, while the latter uses the best worker and memory settings found for each platform/backend pair.
For this reason, the figure should be read as an end-to-end result for the full methodology rather than as a policy-only comparison.

\subsection{Performance Portability}
\label{sec:exp_pp_view}
\begin{figure}[t]
\centering
\includestandalone[width=0.78\linewidth]{pictures/tikz/pp_app}
\caption{Application efficiency and performance-portability view of the tuned methodology. Each matrix cell reports application efficiency relative to the best backend on the same platform; gray indicates unsupported platform/backend combinations. The lower rows report the harmonic and arithmetic means across the investigated platforms.}
\label{fig:pp-app}
\end{figure}

This experiment summarizes how the tuned methodology transfers across the supported platforms and backends.
Here, the tuned methodology is the same \emph{Tuned} configuration introduced in the previous subsection.
For each supported platform/backend pair, we consider the best configuration selected from the worker/memory study, that is, the worker count and per-worker memory budget used for the Tuned bars in Figure~\ref{fig:headline_baseline_tuned}.
To evaluate portability we use two complementary views.
First, we report application throughput in ligands per second, which is the primary metric for end-to-end screening.
Second, we discuss application efficiency, defined as the throughput of one backend on a platform divided by the best backend throughput observed on that same platform.
This distinction is important: absolute throughput captures practical usefulness, while efficiency highlights how close each programming model comes to the best available result on the target machine.

Figure~\ref{fig:pp-app} reports application efficiency, defined as throughput normalized to the best Tuned backend on the same platform.
The lower rows report two summaries across platforms: the harmonic mean and the arithmetic mean of efficiency.
We use the harmonic mean as the stricter portability-oriented metric, because it penalizes weak or unsupported combinations more strongly, while the arithmetic mean is reported as a complementary average-efficiency view~\cite{pennycook2019implications,pennycook2021revisiting}.

The results show that the tuned methodology retains a good platform-local efficiency while exposing the remaining portability limits across backends.
This is the trend shown by \cref{fig:pp-app}:
The matrix reports high local efficiency on the supported backends, while the summary rows show the portability cost of unsupported combinations and backend-dependent performance differences.

%% file: pictures/tikz/bucket_mult_gh200_cuda.tex
\begin{tikzpicture}
\begin{groupplot}[
  group style={
    group size=2 by 1,
    horizontal sep=1cm,
    ylabels at=edge left,
    xlabels at=edge bottom
  },
  width=0.44\textwidth,
  height=0.44\textwidth,
  xmin=1, xmax=7,
  xtick={1,2,3,4,5,6,7},
  xticklabels={1,2,3,4,5,6,7},
  xticklabel style={/pgf/number format/fixed},
  ymin=0, ymax=7000,
  ytick={0,2000,4000,6000},
  xlabel={Memory Budget},
  x unit={\unit{\giga\byte}},
  ylabel={Throughput},
  y unit={ligands/\unit{\second}},
  grid=both,
  grid style={draw=gray!20},
  major grid style={draw=gray!30},
  unbounded coords=jump,
  tick label style={font=\footnotesize},
  label style={font=\footnotesize},
  title style={font=\footnotesize},
  legend style={font=\footnotesize, draw=none, fill=none, column sep=8pt},
  every axis plot/.append style={line width=1.1pt, mark size=2.2pt},
]

\nextgroupplot[
  title={Concurrent GPU Access},
  legend to name=memworklegendgh200cuda,
  legend columns=3,
]

\addplot[name path=up1, draw=none, forget plot] table [
  x index=0,
  y expr=\thisrowno{2}+\thisrowno{3},
  restrict expr to domain={\thisrowno{1}}{1:1}
] {pictures/data/bucket-mult/gh200-cuda/throughput_vs_bucket_gh200_cuda_kernel_lock_off.dat};

\addplot[name path=down1, draw=none, forget plot] table [
  x index=0,
  y expr=\thisrowno{2}-\thisrowno{3},
  restrict expr to domain={\thisrowno{1}}{1:1}
] {pictures/data/bucket-mult/gh200-cuda/throughput_vs_bucket_gh200_cuda_kernel_lock_off.dat};

\addplot[blue!80!black, fill opacity=0.18, forget plot] fill between[of=up1 and down1];

\addplot+[color=blue!80!black, solid, mark=*] table [
  x index=0,
  y index=2,
  restrict expr to domain={\thisrowno{1}}{1:1}
] {pictures/data/bucket-mult/gh200-cuda/throughput_vs_bucket_gh200_cuda_kernel_lock_off.dat};
\addlegendentry{$W=1$}

\addplot[name path=up2, draw=none, forget plot] table [
  x index=0,
  y expr=\thisrowno{2}+\thisrowno{3},
  restrict expr to domain={\thisrowno{1}}{2:2}
] {pictures/data/bucket-mult/gh200-cuda/throughput_vs_bucket_gh200_cuda_kernel_lock_off.dat};

\addplot[name path=down2, draw=none, forget plot] table [
  x index=0,
  y expr=\thisrowno{2}-\thisrowno{3},
  restrict expr to domain={\thisrowno{1}}{2:2}
] {pictures/data/bucket-mult/gh200-cuda/throughput_vs_bucket_gh200_cuda_kernel_lock_off.dat};

\addplot[teal!80!black, fill opacity=0.18, forget plot] fill between[of=up2 and down2];

\addplot+[color=teal!80!black, solid, mark=square*] table [
  x index=0,
  y index=2,
  restrict expr to domain={\thisrowno{1}}{2:2}
] {pictures/data/bucket-mult/gh200-cuda/throughput_vs_bucket_gh200_cuda_kernel_lock_off.dat};
\addlegendentry{$W=2$}

\addplot[name path=up4, draw=none, forget plot] table [
  x index=0,
  y expr=\thisrowno{2}+\thisrowno{3},
  restrict expr to domain={\thisrowno{1}}{4:4}
] {pictures/data/bucket-mult/gh200-cuda/throughput_vs_bucket_gh200_cuda_kernel_lock_off.dat};

\addplot[name path=down4, draw=none, forget plot] table [
  x index=0,
  y expr=\thisrowno{2}-\thisrowno{3},
  restrict expr to domain={\thisrowno{1}}{4:4}
] {pictures/data/bucket-mult/gh200-cuda/throughput_vs_bucket_gh200_cuda_kernel_lock_off.dat};

\addplot[green!50!black, fill opacity=0.18, forget plot] fill between[of=up4 and down4];

\addplot+[color=green!50!black, solid, mark=triangle*] table [
  x index=0,
  y index=2,
  restrict expr to domain={\thisrowno{1}}{4:4}
] {pictures/data/bucket-mult/gh200-cuda/throughput_vs_bucket_gh200_cuda_kernel_lock_off.dat};
\addlegendentry{$W=4$}

\addplot[name path=up8, draw=none, forget plot] table [
  x index=0,
  y expr=\thisrowno{2}+\thisrowno{3},
  restrict expr to domain={\thisrowno{1}}{8:8}
] {pictures/data/bucket-mult/gh200-cuda/throughput_vs_bucket_gh200_cuda_kernel_lock_off.dat};

\addplot[name path=down8, draw=none, forget plot] table [
  x index=0,
  y expr=\thisrowno{2}-\thisrowno{3},
  restrict expr to domain={\thisrowno{1}}{8:8}
] {pictures/data/bucket-mult/gh200-cuda/throughput_vs_bucket_gh200_cuda_kernel_lock_off.dat};

\addplot[orange!90!black, fill opacity=0.18, forget plot] fill between[of=up8 and down8];

\addplot+[color=orange!90!black, solid, mark=diamond*] table [
  x index=0,
  y index=2,
  restrict expr to domain={\thisrowno{1}}{8:8}
] {pictures/data/bucket-mult/gh200-cuda/throughput_vs_bucket_gh200_cuda_kernel_lock_off.dat};
\addlegendentry{$W=8$}

\addplot[name path=up16, draw=none, forget plot] table [
  x index=0,
  y expr=\thisrowno{2}+\thisrowno{3},
  restrict expr to domain={\thisrowno{1}}{16:16}
] {pictures/data/bucket-mult/gh200-cuda/throughput_vs_bucket_gh200_cuda_kernel_lock_off.dat};

\addplot[name path=down16, draw=none, forget plot] table [
  x index=0,
  y expr=\thisrowno{2}-\thisrowno{3},
  restrict expr to domain={\thisrowno{1}}{16:16}
] {pictures/data/bucket-mult/gh200-cuda/throughput_vs_bucket_gh200_cuda_kernel_lock_off.dat};

\addplot[magenta!80!black, fill opacity=0.18, forget plot] fill between[of=up16 and down16];

\addplot+[color=magenta!80!black, solid, mark=pentagon*] table [
  x index=0,
  y index=2,
  restrict expr to domain={\thisrowno{1}}{16:16}
] {pictures/data/bucket-mult/gh200-cuda/throughput_vs_bucket_gh200_cuda_kernel_lock_off.dat};
\addlegendentry{$W=16$}

\nextgroupplot[
  title={Locked GPU Access},
  ylabel={},
]

\addplot[name path=mbup1, draw=none] table [
  x index=0,
  y expr=\thisrowno{2}+\thisrowno{3},
  restrict expr to domain={\thisrowno{1}}{1:1}
] {pictures/data/bucket-mult/gh200-cuda/throughput_vs_bucket_gh200_cuda_kernel_lock_on.dat};

\addplot[name path=mbdown1, draw=none] table [
  x index=0,
  y expr=\thisrowno{2}-\thisrowno{3},
  restrict expr to domain={\thisrowno{1}}{1:1}
] {pictures/data/bucket-mult/gh200-cuda/throughput_vs_bucket_gh200_cuda_kernel_lock_on.dat};

\addplot[blue!80!black, fill opacity=0.18] fill between[of=mbup1 and mbdown1];

\addplot+[color=blue!80!black, solid, mark=*] table [
  x index=0,
  y index=2,
  restrict expr to domain={\thisrowno{1}}{1:1}
] {pictures/data/bucket-mult/gh200-cuda/throughput_vs_bucket_gh200_cuda_kernel_lock_on.dat};

\addplot[name path=mbup2, draw=none] table [
  x index=0,
  y expr=\thisrowno{2}+\thisrowno{3},
  restrict expr to domain={\thisrowno{1}}{2:2}
] {pictures/data/bucket-mult/gh200-cuda/throughput_vs_bucket_gh200_cuda_kernel_lock_on.dat};

\addplot[name path=mbdown2, draw=none] table [
  x index=0,
  y expr=\thisrowno{2}-\thisrowno{3},
  restrict expr to domain={\thisrowno{1}}{2:2}
] {pictures/data/bucket-mult/gh200-cuda/throughput_vs_bucket_gh200_cuda_kernel_lock_on.dat};

\addplot[teal!80!black, fill opacity=0.18] fill between[of=mbup2 and mbdown2];

\addplot+[color=teal!80!black, solid, mark=square*] table [
  x index=0,
  y index=2,
  restrict expr to domain={\thisrowno{1}}{2:2}
] {pictures/data/bucket-mult/gh200-cuda/throughput_vs_bucket_gh200_cuda_kernel_lock_on.dat};

\addplot[name path=mbup4, draw=none] table [
  x index=0,
  y expr=\thisrowno{2}+\thisrowno{3},
  restrict expr to domain={\thisrowno{1}}{4:4}
] {pictures/data/bucket-mult/gh200-cuda/throughput_vs_bucket_gh200_cuda_kernel_lock_on.dat};

\addplot[name path=mbdown4, draw=none] table [
  x index=0,
  y expr=\thisrowno{2}-\thisrowno{3},
  restrict expr to domain={\thisrowno{1}}{4:4}
] {pictures/data/bucket-mult/gh200-cuda/throughput_vs_bucket_gh200_cuda_kernel_lock_on.dat};

\addplot[green!50!black, fill opacity=0.18] fill between[of=mbup4 and mbdown4];

\addplot+[color=green!50!black, solid, mark=triangle*] table [
  x index=0,
  y index=2,
  restrict expr to domain={\thisrowno{1}}{4:4}
] {pictures/data/bucket-mult/gh200-cuda/throughput_vs_bucket_gh200_cuda_kernel_lock_on.dat};

\addplot[name path=mbup8, draw=none] table [
  x index=0,
  y expr=\thisrowno{2}+\thisrowno{3},
  restrict expr to domain={\thisrowno{1}}{8:8}
] {pictures/data/bucket-mult/gh200-cuda/throughput_vs_bucket_gh200_cuda_kernel_lock_on.dat};

\addplot[name path=mbdown8, draw=none] table [
  x index=0,
  y expr=\thisrowno{2}-\thisrowno{3},
  restrict expr to domain={\thisrowno{1}}{8:8}
] {pictures/data/bucket-mult/gh200-cuda/throughput_vs_bucket_gh200_cuda_kernel_lock_on.dat};

\addplot[orange!90!black, fill opacity=0.18] fill between[of=mbup8 and mbdown8];

\addplot+[color=orange!90!black, solid, mark=diamond*] table [
  x index=0,
  y index=2,
  restrict expr to domain={\thisrowno{1}}{8:8}
] {pictures/data/bucket-mult/gh200-cuda/throughput_vs_bucket_gh200_cuda_kernel_lock_on.dat};

\addplot[name path=mbup16, draw=none] table [
  x index=0,
  y expr=\thisrowno{2}+\thisrowno{3},
  restrict expr to domain={\thisrowno{1}}{16:16}
] {pictures/data/bucket-mult/gh200-cuda/throughput_vs_bucket_gh200_cuda_kernel_lock_on.dat};

\addplot[name path=mbdown16, draw=none] table [
  x index=0,
  y expr=\thisrowno{2}-\thisrowno{3},
  restrict expr to domain={\thisrowno{1}}{16:16}
] {pictures/data/bucket-mult/gh200-cuda/throughput_vs_bucket_gh200_cuda_kernel_lock_on.dat};

\addplot[magenta!80!black, fill opacity=0.18] fill between[of=mbup16 and mbdown16];

\addplot+[color=magenta!80!black, solid, mark=pentagon*] table [
  x index=0,
  y index=2,
  restrict expr to domain={\thisrowno{1}}{16:16}
] {pictures/data/bucket-mult/gh200-cuda/throughput_vs_bucket_gh200_cuda_kernel_lock_on.dat};

\end{groupplot}
\end{tikzpicture}

%% file: pictures/tikz/bucket_mult_gh200_hip.tex
\begin{tikzpicture}
\begin{groupplot}[
  group style={
    group size=2 by 1,
    horizontal sep=1cm,
    ylabels at=edge left,
    xlabels at=edge bottom
  },
  width=0.44\textwidth,
  height=0.44\textwidth,
  xmin=1, xmax=7,
  xtick={1,2,3,4,5,6,7},
  xticklabels={1,2,3,4,5,6,7},
  xticklabel style={/pgf/number format/fixed},
  ymin=0, ymax=7000,
  ytick={0,2000,4000,6000},
  xlabel={Memory Budget},
  x unit={\unit{\giga\byte}},
  ylabel={Throughput},
  y unit={ligands/\unit{\second}},
  grid=both,
  grid style={draw=gray!20},
  major grid style={draw=gray!30},
  unbounded coords=jump,
  tick label style={font=\footnotesize},
  label style={font=\footnotesize},
  title style={font=\footnotesize},
  legend style={font=\footnotesize, draw=none, fill=none, column sep=8pt},
  every axis plot/.append style={line width=1.1pt, mark size=2.2pt},
]

\nextgroupplot[
  title={Concurrent GPU Access},
  legend to name=memworklegendgh200hip,
  legend columns=3,
]

\addplot[name path=up1, draw=none, forget plot] table [
  x index=0,
  y expr=\thisrowno{2}+\thisrowno{3},
  restrict expr to domain={\thisrowno{1}}{1:1}
] {pictures/data/bucket-mult/gh200-hip/throughput_vs_bucket_gh200_hip_kernel_lock_off.dat};

\addplot[name path=down1, draw=none, forget plot] table [
  x index=0,
  y expr=\thisrowno{2}-\thisrowno{3},
  restrict expr to domain={\thisrowno{1}}{1:1}
] {pictures/data/bucket-mult/gh200-hip/throughput_vs_bucket_gh200_hip_kernel_lock_off.dat};

\addplot[blue!80!black, fill opacity=0.18, forget plot] fill between[of=up1 and down1];

\addplot+[color=blue!80!black, solid, mark=*] table [
  x index=0,
  y index=2,
  restrict expr to domain={\thisrowno{1}}{1:1}
] {pictures/data/bucket-mult/gh200-hip/throughput_vs_bucket_gh200_hip_kernel_lock_off.dat};
\addlegendentry{$W=1$}

\addplot[name path=up2, draw=none, forget plot] table [
  x index=0,
  y expr=\thisrowno{2}+\thisrowno{3},
  restrict expr to domain={\thisrowno{1}}{2:2}
] {pictures/data/bucket-mult/gh200-hip/throughput_vs_bucket_gh200_hip_kernel_lock_off.dat};

\addplot[name path=down2, draw=none, forget plot] table [
  x index=0,
  y expr=\thisrowno{2}-\thisrowno{3},
  restrict expr to domain={\thisrowno{1}}{2:2}
] {pictures/data/bucket-mult/gh200-hip/throughput_vs_bucket_gh200_hip_kernel_lock_off.dat};

\addplot[teal!80!black, fill opacity=0.18, forget plot] fill between[of=up2 and down2];

\addplot+[color=teal!80!black, solid, mark=square*] table [
  x index=0,
  y index=2,
  restrict expr to domain={\thisrowno{1}}{2:2}
] {pictures/data/bucket-mult/gh200-hip/throughput_vs_bucket_gh200_hip_kernel_lock_off.dat};
\addlegendentry{$W=2$}

\addplot[name path=up4, draw=none, forget plot] table [
  x index=0,
  y expr=\thisrowno{2}+\thisrowno{3},
  restrict expr to domain={\thisrowno{1}}{4:4}
] {pictures/data/bucket-mult/gh200-hip/throughput_vs_bucket_gh200_hip_kernel_lock_off.dat};

\addplot[name path=down4, draw=none, forget plot] table [
  x index=0,
  y expr=\thisrowno{2}-\thisrowno{3},
  restrict expr to domain={\thisrowno{1}}{4:4}
] {pictures/data/bucket-mult/gh200-hip/throughput_vs_bucket_gh200_hip_kernel_lock_off.dat};

\addplot[green!50!black, fill opacity=0.18, forget plot] fill between[of=up4 and down4];

\addplot+[color=green!50!black, solid, mark=triangle*] table [
  x index=0,
  y index=2,
  restrict expr to domain={\thisrowno{1}}{4:4}
] {pictures/data/bucket-mult/gh200-hip/throughput_vs_bucket_gh200_hip_kernel_lock_off.dat};
\addlegendentry{$W=4$}

\addplot[name path=up8, draw=none, forget plot] table [
  x index=0,
  y expr=\thisrowno{2}+\thisrowno{3},
  restrict expr to domain={\thisrowno{1}}{8:8}
] {pictures/data/bucket-mult/gh200-hip/throughput_vs_bucket_gh200_hip_kernel_lock_off.dat};

\addplot[name path=down8, draw=none, forget plot] table [
  x index=0,
  y expr=\thisrowno{2}-\thisrowno{3},
  restrict expr to domain={\thisrowno{1}}{8:8}
] {pictures/data/bucket-mult/gh200-hip/throughput_vs_bucket_gh200_hip_kernel_lock_off.dat};

\addplot[orange!90!black, fill opacity=0.18, forget plot] fill between[of=up8 and down8];

\addplot+[color=orange!90!black, solid, mark=diamond*] table [
  x index=0,
  y index=2,
  restrict expr to domain={\thisrowno{1}}{8:8}
] {pictures/data/bucket-mult/gh200-hip/throughput_vs_bucket_gh200_hip_kernel_lock_off.dat};
\addlegendentry{$W=8$}

\addplot[name path=up16, draw=none, forget plot] table [
  x index=0,
  y expr=\thisrowno{2}+\thisrowno{3},
  restrict expr to domain={\thisrowno{1}}{16:16}
] {pictures/data/bucket-mult/gh200-hip/throughput_vs_bucket_gh200_hip_kernel_lock_off.dat};

\addplot[name path=down16, draw=none, forget plot] table [
  x index=0,
  y expr=\thisrowno{2}-\thisrowno{3},
  restrict expr to domain={\thisrowno{1}}{16:16}
] {pictures/data/bucket-mult/gh200-hip/throughput_vs_bucket_gh200_hip_kernel_lock_off.dat};

\addplot[magenta!80!black, fill opacity=0.18, forget plot] fill between[of=up16 and down16];

\addplot+[color=magenta!80!black, solid, mark=pentagon*] table [
  x index=0,
  y index=2,
  restrict expr to domain={\thisrowno{1}}{16:16}
] {pictures/data/bucket-mult/gh200-hip/throughput_vs_bucket_gh200_hip_kernel_lock_off.dat};
\addlegendentry{$W=16$}

\nextgroupplot[
  title={Locked GPU Access},
  ylabel={},
]

\addplot[name path=mbup1, draw=none] table [
  x index=0,
  y expr=\thisrowno{2}+\thisrowno{3},
  restrict expr to domain={\thisrowno{1}}{1:1}
] {pictures/data/bucket-mult/gh200-hip/throughput_vs_bucket_gh200_hip_kernel_lock_on.dat};

\addplot[name path=mbdown1, draw=none] table [
  x index=0,
  y expr=\thisrowno{2}-\thisrowno{3},
  restrict expr to domain={\thisrowno{1}}{1:1}
] {pictures/data/bucket-mult/gh200-hip/throughput_vs_bucket_gh200_hip_kernel_lock_on.dat};

\addplot[blue!80!black, fill opacity=0.18] fill between[of=mbup1 and mbdown1];

\addplot+[color=blue!80!black, solid, mark=*] table [
  x index=0,
  y index=2,
  restrict expr to domain={\thisrowno{1}}{1:1}
] {pictures/data/bucket-mult/gh200-hip/throughput_vs_bucket_gh200_hip_kernel_lock_on.dat};

\addplot[name path=mbup2, draw=none] table [
  x index=0,
  y expr=\thisrowno{2}+\thisrowno{3},
  restrict expr to domain={\thisrowno{1}}{2:2}
] {pictures/data/bucket-mult/gh200-hip/throughput_vs_bucket_gh200_hip_kernel_lock_on.dat};

\addplot[name path=mbdown2, draw=none] table [
  x index=0,
  y expr=\thisrowno{2}-\thisrowno{3},
  restrict expr to domain={\thisrowno{1}}{2:2}
] {pictures/data/bucket-mult/gh200-hip/throughput_vs_bucket_gh200_hip_kernel_lock_on.dat};

\addplot[teal!80!black, fill opacity=0.18] fill between[of=mbup2 and mbdown2];

\addplot+[color=teal!80!black, solid, mark=square*] table [
  x index=0,
  y index=2,
  restrict expr to domain={\thisrowno{1}}{2:2}
] {pictures/data/bucket-mult/gh200-hip/throughput_vs_bucket_gh200_hip_kernel_lock_on.dat};

\addplot[name path=mbup4, draw=none] table [
  x index=0,
  y expr=\thisrowno{2}+\thisrowno{3},
  restrict expr to domain={\thisrowno{1}}{4:4}
] {pictures/data/bucket-mult/gh200-hip/throughput_vs_bucket_gh200_hip_kernel_lock_on.dat};

\addplot[name path=mbdown4, draw=none] table [
  x index=0,
  y expr=\thisrowno{2}-\thisrowno{3},
  restrict expr to domain={\thisrowno{1}}{4:4}
] {pictures/data/bucket-mult/gh200-hip/throughput_vs_bucket_gh200_hip_kernel_lock_on.dat};

\addplot[green!50!black, fill opacity=0.18] fill between[of=mbup4 and mbdown4];

\addplot+[color=green!50!black, solid, mark=triangle*] table [
  x index=0,
  y index=2,
  restrict expr to domain={\thisrowno{1}}{4:4}
] {pictures/data/bucket-mult/gh200-hip/throughput_vs_bucket_gh200_hip_kernel_lock_on.dat};

\addplot[name path=mbup8, draw=none] table [
  x index=0,
  y expr=\thisrowno{2}+\thisrowno{3},
  restrict expr to domain={\thisrowno{1}}{8:8}
] {pictures/data/bucket-mult/gh200-hip/throughput_vs_bucket_gh200_hip_kernel_lock_on.dat};

\addplot[name path=mbdown8, draw=none] table [
  x index=0,
  y expr=\thisrowno{2}-\thisrowno{3},
  restrict expr to domain={\thisrowno{1}}{8:8}
] {pictures/data/bucket-mult/gh200-hip/throughput_vs_bucket_gh200_hip_kernel_lock_on.dat};

\addplot[orange!90!black, fill opacity=0.18] fill between[of=mbup8 and mbdown8];

\addplot+[color=orange!90!black, solid, mark=diamond*] table [
  x index=0,
  y index=2,
  restrict expr to domain={\thisrowno{1}}{8:8}
] {pictures/data/bucket-mult/gh200-hip/throughput_vs_bucket_gh200_hip_kernel_lock_on.dat};

\addplot[name path=mbup16, draw=none] table [
  x index=0,
  y expr=\thisrowno{2}+\thisrowno{3},
  restrict expr to domain={\thisrowno{1}}{16:16}
] {pictures/data/bucket-mult/gh200-hip/throughput_vs_bucket_gh200_hip_kernel_lock_on.dat};

\addplot[name path=mbdown16, draw=none] table [
  x index=0,
  y expr=\thisrowno{2}-\thisrowno{3},
  restrict expr to domain={\thisrowno{1}}{16:16}
] {pictures/data/bucket-mult/gh200-hip/throughput_vs_bucket_gh200_hip_kernel_lock_on.dat};

\addplot[magenta!80!black, fill opacity=0.18] fill between[of=mbup16 and mbdown16];

\addplot+[color=magenta!80!black, solid, mark=pentagon*] table [
  x index=0,
  y index=2,
  restrict expr to domain={\thisrowno{1}}{16:16}
] {pictures/data/bucket-mult/gh200-hip/throughput_vs_bucket_gh200_hip_kernel_lock_on.dat};

\end{groupplot}
\end{tikzpicture}

%% file: pictures/tikz/bucket_mult_gh200_sycl.tex
\begin{tikzpicture}
\begin{groupplot}[
  group style={
    group size=2 by 1,
    horizontal sep=1cm,
    ylabels at=edge left,
    xlabels at=edge bottom
  },
  width=0.44\textwidth,
  height=0.44\textwidth,
  xmin=1, xmax=7,
  xtick={1,2,3,4,5,6,7},
  xticklabels={1,2,3,4,5,6,7},
  xticklabel style={/pgf/number format/fixed},
  ymin=0, ymax=4000,
  ytick={0,1000,2000,3000, 4000},
  xlabel={Memory Budget},
  x unit={\unit{\giga\byte}},
  ylabel={Throughput},
  y unit={ligands/\unit{\second}},
  grid=both,
  grid style={draw=gray!20},
  major grid style={draw=gray!30},
  unbounded coords=jump,
  tick label style={font=\footnotesize},
  label style={font=\footnotesize},
  title style={font=\footnotesize},
  legend style={font=\footnotesize, draw=none, fill=none, column sep=8pt},
  every axis plot/.append style={line width=1.1pt, mark size=2.2pt},
]

\nextgroupplot[
  legend to name=memworklegendgh200sycl,
  legend columns=3,
]

\addplot[name path=up1, draw=none, forget plot] table [
  x index=0,
  y expr=\thisrowno{2}+\thisrowno{3},
  restrict expr to domain={\thisrowno{1}}{1:1}
] {pictures/data/bucket-mult/gh200-sycl/throughput_vs_bucket_gh200_sycl_kernel_lock_off.dat};

\addplot[name path=down1, draw=none, forget plot] table [
  x index=0,
  y expr=\thisrowno{2}-\thisrowno{3},
  restrict expr to domain={\thisrowno{1}}{1:1}
] {pictures/data/bucket-mult/gh200-sycl/throughput_vs_bucket_gh200_sycl_kernel_lock_off.dat};

\addplot[blue!80!black, fill opacity=0.18, forget plot] fill between[of=up1 and down1];

\addplot+[color=blue!80!black, solid, mark=*] table [
  x index=0,
  y index=2,
  restrict expr to domain={\thisrowno{1}}{1:1}
] {pictures/data/bucket-mult/gh200-sycl/throughput_vs_bucket_gh200_sycl_kernel_lock_off.dat};
\addlegendentry{$W=1$}

\addplot[name path=up2, draw=none, forget plot] table [
  x index=0,
  y expr=\thisrowno{2}+\thisrowno{3},
  restrict expr to domain={\thisrowno{1}}{2:2}
] {pictures/data/bucket-mult/gh200-sycl/throughput_vs_bucket_gh200_sycl_kernel_lock_off.dat};

\addplot[name path=down2, draw=none, forget plot] table [
  x index=0,
  y expr=\thisrowno{2}-\thisrowno{3},
  restrict expr to domain={\thisrowno{1}}{2:2}
] {pictures/data/bucket-mult/gh200-sycl/throughput_vs_bucket_gh200_sycl_kernel_lock_off.dat};

\addplot[teal!80!black, fill opacity=0.18, forget plot] fill between[of=up2 and down2];

\addplot+[color=teal!80!black, solid, mark=square*] table [
  x index=0,
  y index=2,
  restrict expr to domain={\thisrowno{1}}{2:2}
] {pictures/data/bucket-mult/gh200-sycl/throughput_vs_bucket_gh200_sycl_kernel_lock_off.dat};
\addlegendentry{$W=2$}

\addplot[name path=up4, draw=none, forget plot] table [
  x index=0,
  y expr=\thisrowno{2}+\thisrowno{3},
  restrict expr to domain={\thisrowno{1}}{4:4}
] {pictures/data/bucket-mult/gh200-sycl/throughput_vs_bucket_gh200_sycl_kernel_lock_off.dat};

\addplot[name path=down4, draw=none, forget plot] table [
  x index=0,
  y expr=\thisrowno{2}-\thisrowno{3},
  restrict expr to domain={\thisrowno{1}}{4:4}
] {pictures/data/bucket-mult/gh200-sycl/throughput_vs_bucket_gh200_sycl_kernel_lock_off.dat};

\addplot[green!50!black, fill opacity=0.18, forget plot] fill between[of=up4 and down4];

\addplot+[color=green!50!black, solid, mark=triangle*] table [
  x index=0,
  y index=2,
  restrict expr to domain={\thisrowno{1}}{4:4}
] {pictures/data/bucket-mult/gh200-sycl/throughput_vs_bucket_gh200_sycl_kernel_lock_off.dat};
\addlegendentry{$W=4$}

\addplot[name path=up8, draw=none, forget plot] table [
  x index=0,
  y expr=\thisrowno{2}+\thisrowno{3},
  restrict expr to domain={\thisrowno{1}}{8:8}
] {pictures/data/bucket-mult/gh200-sycl/throughput_vs_bucket_gh200_sycl_kernel_lock_off.dat};

\addplot[name path=down8, draw=none, forget plot] table [
  x index=0,
  y expr=\thisrowno{2}-\thisrowno{3},
  restrict expr to domain={\thisrowno{1}}{8:8}
] {pictures/data/bucket-mult/gh200-sycl/throughput_vs_bucket_gh200_sycl_kernel_lock_off.dat};

\addplot[orange!90!black, fill opacity=0.18, forget plot] fill between[of=up8 and down8];

\addplot+[color=orange!90!black, solid, mark=diamond*] table [
  x index=0,
  y index=2,
  restrict expr to domain={\thisrowno{1}}{8:8}
] {pictures/data/bucket-mult/gh200-sycl/throughput_vs_bucket_gh200_sycl_kernel_lock_off.dat};
\addlegendentry{$W=8$}

\addplot[name path=up16, draw=none, forget plot] table [
  x index=0,
  y expr=\thisrowno{2}+\thisrowno{3},
  restrict expr to domain={\thisrowno{1}}{16:16}
] {pictures/data/bucket-mult/gh200-sycl/throughput_vs_bucket_gh200_sycl_kernel_lock_off.dat};

\addplot[name path=down16, draw=none, forget plot] table [
  x index=0,
  y expr=\thisrowno{2}-\thisrowno{3},
  restrict expr to domain={\thisrowno{1}}{16:16}
] {pictures/data/bucket-mult/gh200-sycl/throughput_vs_bucket_gh200_sycl_kernel_lock_off.dat};

\addplot[magenta!80!black, fill opacity=0.18, forget plot] fill between[of=up16 and down16];

\addplot+[color=magenta!80!black, solid, mark=pentagon*] table [
  x index=0,
  y index=2,
  restrict expr to domain={\thisrowno{1}}{16:16}
] {pictures/data/bucket-mult/gh200-sycl/throughput_vs_bucket_gh200_sycl_kernel_lock_off.dat};
\addlegendentry{$W=16$}

\nextgroupplot[
  ylabel={},
]

\addplot[name path=mbup1, draw=none] table [
  x index=0,
  y expr=\thisrowno{2}+\thisrowno{3},
  restrict expr to domain={\thisrowno{1}}{1:1}
] {pictures/data/bucket-mult/gh200-sycl/throughput_vs_bucket_gh200_sycl_kernel_lock_on.dat};

\addplot[name path=mbdown1, draw=none] table [
  x index=0,
  y expr=\thisrowno{2}-\thisrowno{3},
  restrict expr to domain={\thisrowno{1}}{1:1}
] {pictures/data/bucket-mult/gh200-sycl/throughput_vs_bucket_gh200_sycl_kernel_lock_on.dat};

\addplot[blue!80!black, fill opacity=0.18] fill between[of=mbup1 and mbdown1];

\addplot+[color=blue!80!black, solid, mark=*] table [
  x index=0,
  y index=2,
  restrict expr to domain={\thisrowno{1}}{1:1}
] {pictures/data/bucket-mult/gh200-sycl/throughput_vs_bucket_gh200_sycl_kernel_lock_on.dat};

\addplot[name path=mbup2, draw=none] table [
  x index=0,
  y expr=\thisrowno{2}+\thisrowno{3},
  restrict expr to domain={\thisrowno{1}}{2:2}
] {pictures/data/bucket-mult/gh200-sycl/throughput_vs_bucket_gh200_sycl_kernel_lock_on.dat};

\addplot[name path=mbdown2, draw=none] table [
  x index=0,
  y expr=\thisrowno{2}-\thisrowno{3},
  restrict expr to domain={\thisrowno{1}}{2:2}
] {pictures/data/bucket-mult/gh200-sycl/throughput_vs_bucket_gh200_sycl_kernel_lock_on.dat};

\addplot[teal!80!black, fill opacity=0.18] fill between[of=mbup2 and mbdown2];

\addplot+[color=teal!80!black, solid, mark=square*] table [
  x index=0,
  y index=2,
  restrict expr to domain={\thisrowno{1}}{2:2}
] {pictures/data/bucket-mult/gh200-sycl/throughput_vs_bucket_gh200_sycl_kernel_lock_on.dat};

\addplot[name path=mbup4, draw=none] table [
  x index=0,
  y expr=\thisrowno{2}+\thisrowno{3},
  restrict expr to domain={\thisrowno{1}}{4:4}
] {pictures/data/bucket-mult/gh200-sycl/throughput_vs_bucket_gh200_sycl_kernel_lock_on.dat};

\addplot[name path=mbdown4, draw=none] table [
  x index=0,
  y expr=\thisrowno{2}-\thisrowno{3},
  restrict expr to domain={\thisrowno{1}}{4:4}
] {pictures/data/bucket-mult/gh200-sycl/throughput_vs_bucket_gh200_sycl_kernel_lock_on.dat};

\addplot[green!50!black, fill opacity=0.18] fill between[of=mbup4 and mbdown4];

\addplot+[color=green!50!black, solid, mark=triangle*] table [
  x index=0,
  y index=2,
  restrict expr to domain={\thisrowno{1}}{4:4}
] {pictures/data/bucket-mult/gh200-sycl/throughput_vs_bucket_gh200_sycl_kernel_lock_on.dat};

\addplot[name path=mbup8, draw=none] table [
  x index=0,
  y expr=\thisrowno{2}+\thisrowno{3},
  restrict expr to domain={\thisrowno{1}}{8:8}
] {pictures/data/bucket-mult/gh200-sycl/throughput_vs_bucket_gh200_sycl_kernel_lock_on.dat};

\addplot[name path=mbdown8, draw=none] table [
  x index=0,
  y expr=\thisrowno{2}-\thisrowno{3},
  restrict expr to domain={\thisrowno{1}}{8:8}
] {pictures/data/bucket-mult/gh200-sycl/throughput_vs_bucket_gh200_sycl_kernel_lock_on.dat};

\addplot[orange!90!black, fill opacity=0.18] fill between[of=mbup8 and mbdown8];

\addplot+[color=orange!90!black, solid, mark=diamond*] table [
  x index=0,
  y index=2,
  restrict expr to domain={\thisrowno{1}}{8:8}
] {pictures/data/bucket-mult/gh200-sycl/throughput_vs_bucket_gh200_sycl_kernel_lock_on.dat};

\addplot[name path=mbup16, draw=none] table [
  x index=0,
  y expr=\thisrowno{2}+\thisrowno{3},
  restrict expr to domain={\thisrowno{1}}{16:16}
] {pictures/data/bucket-mult/gh200-sycl/throughput_vs_bucket_gh200_sycl_kernel_lock_on.dat};

\addplot[name path=mbdown16, draw=none] table [
  x index=0,
  y expr=\thisrowno{2}-\thisrowno{3},
  restrict expr to domain={\thisrowno{1}}{16:16}
] {pictures/data/bucket-mult/gh200-sycl/throughput_vs_bucket_gh200_sycl_kernel_lock_on.dat};

\addplot[magenta!80!black, fill opacity=0.18] fill between[of=mbup16 and mbdown16];

\addplot+[color=magenta!80!black, solid, mark=pentagon*] table [
  x index=0,
  y index=2,
  restrict expr to domain={\thisrowno{1}}{16:16}
] {pictures/data/bucket-mult/gh200-sycl/throughput_vs_bucket_gh200_sycl_kernel_lock_on.dat};

\end{groupplot}
\end{tikzpicture}

%% file: pictures/tikz/bucket_mult_mi250_hip.tex
\begin{tikzpicture}
\begin{groupplot}[
  group style={
    group size=2 by 1,
    horizontal sep=1cm,
    ylabels at=edge left,
    xlabels at=edge bottom
  },
  width=0.44\textwidth,
  height=0.44\textwidth,
  xmin=1, xmax=7,
  xtick={1,2,3,4,5,6,7},
  xticklabels={1,2,3,4,5,6,7},
  xticklabel style={/pgf/number format/fixed},
  ymin=0, ymax=2000,
  ytick={0,500,1000,1500, 2000},
  xlabel={Memory Budget},
  x unit={\unit{\giga\byte}},
  ylabel={Throughput},
  y unit={ligands/\unit{\second}},
  grid=both,
  grid style={draw=gray!20},
  major grid style={draw=gray!30},
  unbounded coords=jump,
  tick label style={font=\footnotesize},
  label style={font=\footnotesize},
  title style={font=\footnotesize},
  legend style={font=\footnotesize, draw=none, fill=none, column sep=8pt},
  every axis plot/.append style={line width=1.1pt, mark size=2.2pt},
]

\nextgroupplot[
  legend to name=memworklegendmi250hip,
  legend columns=3,
]

\addplot[name path=up1, draw=none, forget plot] table [
  x index=0,
  y expr=\thisrowno{2}+\thisrowno{3},
  restrict expr to domain={\thisrowno{1}}{1:1}
] {pictures/data/bucket-mult/mi250-hip/throughput_vs_bucket_mi250_hip_kernel_lock_off.dat};

\addplot[name path=down1, draw=none, forget plot] table [
  x index=0,
  y expr=\thisrowno{2}-\thisrowno{3},
  restrict expr to domain={\thisrowno{1}}{1:1}
] {pictures/data/bucket-mult/mi250-hip/throughput_vs_bucket_mi250_hip_kernel_lock_off.dat};

\addplot[blue!80!black, fill opacity=0.18, forget plot] fill between[of=up1 and down1];

\addplot+[color=blue!80!black, solid, mark=*] table [
  x index=0,
  y index=2,
  restrict expr to domain={\thisrowno{1}}{1:1}
] {pictures/data/bucket-mult/mi250-hip/throughput_vs_bucket_mi250_hip_kernel_lock_off.dat};
\addlegendentry{$W=1$}

\addplot[name path=up2, draw=none, forget plot] table [
  x index=0,
  y expr=\thisrowno{2}+\thisrowno{3},
  restrict expr to domain={\thisrowno{1}}{2:2}
] {pictures/data/bucket-mult/mi250-hip/throughput_vs_bucket_mi250_hip_kernel_lock_off.dat};

\addplot[name path=down2, draw=none, forget plot] table [
  x index=0,
  y expr=\thisrowno{2}-\thisrowno{3},
  restrict expr to domain={\thisrowno{1}}{2:2}
] {pictures/data/bucket-mult/mi250-hip/throughput_vs_bucket_mi250_hip_kernel_lock_off.dat};

\addplot[teal!80!black, fill opacity=0.18, forget plot] fill between[of=up2 and down2];

\addplot+[color=teal!80!black, solid, mark=square*] table [
  x index=0,
  y index=2,
  restrict expr to domain={\thisrowno{1}}{2:2}
] {pictures/data/bucket-mult/mi250-hip/throughput_vs_bucket_mi250_hip_kernel_lock_off.dat};
\addlegendentry{$W=2$}

\addplot[name path=up4, draw=none, forget plot] table [
  x index=0,
  y expr=\thisrowno{2}+\thisrowno{3},
  restrict expr to domain={\thisrowno{1}}{4:4}
] {pictures/data/bucket-mult/mi250-hip/throughput_vs_bucket_mi250_hip_kernel_lock_off.dat};

\addplot[name path=down4, draw=none, forget plot] table [
  x index=0,
  y expr=\thisrowno{2}-\thisrowno{3},
  restrict expr to domain={\thisrowno{1}}{4:4}
] {pictures/data/bucket-mult/mi250-hip/throughput_vs_bucket_mi250_hip_kernel_lock_off.dat};

\addplot[green!50!black, fill opacity=0.18, forget plot] fill between[of=up4 and down4];

\addplot+[color=green!50!black, solid, mark=triangle*] table [
  x index=0,
  y index=2,
  restrict expr to domain={\thisrowno{1}}{4:4}
] {pictures/data/bucket-mult/mi250-hip/throughput_vs_bucket_mi250_hip_kernel_lock_off.dat};
\addlegendentry{$W=4$}

\addplot[name path=up8, draw=none, forget plot] table [
  x index=0,
  y expr=\thisrowno{2}+\thisrowno{3},
  restrict expr to domain={\thisrowno{1}}{8:8}
] {pictures/data/bucket-mult/mi250-hip/throughput_vs_bucket_mi250_hip_kernel_lock_off.dat};

\addplot[name path=down8, draw=none, forget plot] table [
  x index=0,
  y expr=\thisrowno{2}-\thisrowno{3},
  restrict expr to domain={\thisrowno{1}}{8:8}
] {pictures/data/bucket-mult/mi250-hip/throughput_vs_bucket_mi250_hip_kernel_lock_off.dat};

\addplot[orange!90!black, fill opacity=0.18, forget plot] fill between[of=up8 and down8];

\addplot+[color=orange!90!black, solid, mark=diamond*] table [
  x index=0,
  y index=2,
  restrict expr to domain={\thisrowno{1}}{8:8}
] {pictures/data/bucket-mult/mi250-hip/throughput_vs_bucket_mi250_hip_kernel_lock_off.dat};
\addlegendentry{$W=8$}

\addplot[name path=up16, draw=none, forget plot] table [
  x index=0,
  y expr=\thisrowno{2}+\thisrowno{3},
  restrict expr to domain={\thisrowno{1}}{16:16}
] {pictures/data/bucket-mult/mi250-hip/throughput_vs_bucket_mi250_hip_kernel_lock_off.dat};

\addplot[name path=down16, draw=none, forget plot] table [
  x index=0,
  y expr=\thisrowno{2}-\thisrowno{3},
  restrict expr to domain={\thisrowno{1}}{16:16}
] {pictures/data/bucket-mult/mi250-hip/throughput_vs_bucket_mi250_hip_kernel_lock_off.dat};

\addplot[magenta!80!black, fill opacity=0.18, forget plot] fill between[of=up16 and down16];

\addplot+[color=magenta!80!black, solid, mark=pentagon*] table [
  x index=0,
  y index=2,
  restrict expr to domain={\thisrowno{1}}{16:16}
] {pictures/data/bucket-mult/mi250-hip/throughput_vs_bucket_mi250_hip_kernel_lock_off.dat};
\addlegendentry{$W=16$}

\nextgroupplot[
  ylabel={},
]

\addplot[name path=mbup1, draw=none] table [
  x index=0,
  y expr=\thisrowno{2}+\thisrowno{3},
  restrict expr to domain={\thisrowno{1}}{1:1}
] {pictures/data/bucket-mult/mi250-hip/throughput_vs_bucket_mi250_hip_kernel_lock_on.dat};

\addplot[name path=mbdown1, draw=none] table [
  x index=0,
  y expr=\thisrowno{2}-\thisrowno{3},
  restrict expr to domain={\thisrowno{1}}{1:1}
] {pictures/data/bucket-mult/mi250-hip/throughput_vs_bucket_mi250_hip_kernel_lock_on.dat};

\addplot[blue!80!black, fill opacity=0.18] fill between[of=mbup1 and mbdown1];

\addplot+[color=blue!80!black, solid, mark=*] table [
  x index=0,
  y index=2,
  restrict expr to domain={\thisrowno{1}}{1:1}
] {pictures/data/bucket-mult/mi250-hip/throughput_vs_bucket_mi250_hip_kernel_lock_on.dat};

\addplot[name path=mbup2, draw=none] table [
  x index=0,
  y expr=\thisrowno{2}+\thisrowno{3},
  restrict expr to domain={\thisrowno{1}}{2:2}
] {pictures/data/bucket-mult/mi250-hip/throughput_vs_bucket_mi250_hip_kernel_lock_on.dat};

\addplot[name path=mbdown2, draw=none] table [
  x index=0,
  y expr=\thisrowno{2}-\thisrowno{3},
  restrict expr to domain={\thisrowno{1}}{2:2}
] {pictures/data/bucket-mult/mi250-hip/throughput_vs_bucket_mi250_hip_kernel_lock_on.dat};

\addplot[teal!80!black, fill opacity=0.18] fill between[of=mbup2 and mbdown2];

\addplot+[color=teal!80!black, solid, mark=square*] table [
  x index=0,
  y index=2,
  restrict expr to domain={\thisrowno{1}}{2:2}
] {pictures/data/bucket-mult/mi250-hip/throughput_vs_bucket_mi250_hip_kernel_lock_on.dat};

\addplot[name path=mbup4, draw=none] table [
  x index=0,
  y expr=\thisrowno{2}+\thisrowno{3},
  restrict expr to domain={\thisrowno{1}}{4:4}
] {pictures/data/bucket-mult/mi250-hip/throughput_vs_bucket_mi250_hip_kernel_lock_on.dat};

\addplot[name path=mbdown4, draw=none] table [
  x index=0,
  y expr=\thisrowno{2}-\thisrowno{3},
  restrict expr to domain={\thisrowno{1}}{4:4}
] {pictures/data/bucket-mult/mi250-hip/throughput_vs_bucket_mi250_hip_kernel_lock_on.dat};

\addplot[green!50!black, fill opacity=0.18] fill between[of=mbup4 and mbdown4];

\addplot+[color=green!50!black, solid, mark=triangle*] table [
  x index=0,
  y index=2,
  restrict expr to domain={\thisrowno{1}}{4:4}
] {pictures/data/bucket-mult/mi250-hip/throughput_vs_bucket_mi250_hip_kernel_lock_on.dat};

\addplot[name path=mbup8, draw=none] table [
  x index=0,
  y expr=\thisrowno{2}+\thisrowno{3},
  restrict expr to domain={\thisrowno{1}}{8:8}
] {pictures/data/bucket-mult/mi250-hip/throughput_vs_bucket_mi250_hip_kernel_lock_on.dat};

\addplot[name path=mbdown8, draw=none] table [
  x index=0,
  y expr=\thisrowno{2}-\thisrowno{3},
  restrict expr to domain={\thisrowno{1}}{8:8}
] {pictures/data/bucket-mult/mi250-hip/throughput_vs_bucket_mi250_hip_kernel_lock_on.dat};

\addplot[orange!90!black, fill opacity=0.18] fill between[of=mbup8 and mbdown8];

\addplot+[color=orange!90!black, solid, mark=diamond*] table [
  x index=0,
  y index=2,
  restrict expr to domain={\thisrowno{1}}{8:8}
] {pictures/data/bucket-mult/mi250-hip/throughput_vs_bucket_mi250_hip_kernel_lock_on.dat};

\addplot[name path=mbup16, draw=none] table [
  x index=0,
  y expr=\thisrowno{2}+\thisrowno{3},
  restrict expr to domain={\thisrowno{1}}{16:16}
] {pictures/data/bucket-mult/mi250-hip/throughput_vs_bucket_mi250_hip_kernel_lock_on.dat};

\addplot[name path=mbdown16, draw=none] table [
  x index=0,
  y expr=\thisrowno{2}-\thisrowno{3},
  restrict expr to domain={\thisrowno{1}}{16:16}
] {pictures/data/bucket-mult/mi250-hip/throughput_vs_bucket_mi250_hip_kernel_lock_on.dat};

\addplot[magenta!80!black, fill opacity=0.18] fill between[of=mbup16 and mbdown16];

\addplot+[color=magenta!80!black, solid, mark=pentagon*] table [
  x index=0,
  y index=2,
  restrict expr to domain={\thisrowno{1}}{16:16}
] {pictures/data/bucket-mult/mi250-hip/throughput_vs_bucket_mi250_hip_kernel_lock_on.dat};

\end{groupplot}
\end{tikzpicture}

%% file: pictures/tikz/bucket_mult_mi250_sycl.tex
\begin{tikzpicture}
\begin{groupplot}[
  group style={
    group size=2 by 1,
    horizontal sep=1cm,
    ylabels at=edge left,
    xlabels at=edge bottom
  },
  width=0.44\textwidth,
  height=0.44\textwidth,
  xmin=1, xmax=7,
  xtick={1,2,3,4,5,6,7},
  xticklabels={1,2,3,4,5,6,7},
  xticklabel style={/pgf/number format/fixed},
  ymin=0, ymax=2000,
  ytick={0,500,1000,1500, 2000},
  xlabel={Memory Budget},
  x unit={\unit{\giga\byte}},
  ylabel={Throughput},
  y unit={ligands/\unit{\second}},
  grid=both,
  grid style={draw=gray!20},
  major grid style={draw=gray!30},
  unbounded coords=jump,
  tick label style={font=\footnotesize},
  label style={font=\footnotesize},
  title style={font=\footnotesize},
  legend style={font=\footnotesize, draw=none, fill=none, column sep=8pt},
  every axis plot/.append style={line width=1.1pt, mark size=2.2pt},
]

\nextgroupplot[
  legend to name=memworklegendmi250sycl,
  legend columns=3,
]

\addplot[name path=up1, draw=none, forget plot] table [
  x index=0,
  y expr=\thisrowno{2}+\thisrowno{3},
  restrict expr to domain={\thisrowno{1}}{1:1}
] {pictures/data/bucket-mult/mi250-sycl/throughput_vs_bucket_mi250_sycl_kernel_lock_off.dat};

\addplot[name path=down1, draw=none, forget plot] table [
  x index=0,
  y expr=\thisrowno{2}-\thisrowno{3},
  restrict expr to domain={\thisrowno{1}}{1:1}
] {pictures/data/bucket-mult/mi250-sycl/throughput_vs_bucket_mi250_sycl_kernel_lock_off.dat};

\addplot[blue!80!black, fill opacity=0.18, forget plot] fill between[of=up1 and down1];

\addplot+[color=blue!80!black, solid, mark=*] table [
  x index=0,
  y index=2,
  restrict expr to domain={\thisrowno{1}}{1:1}
] {pictures/data/bucket-mult/mi250-sycl/throughput_vs_bucket_mi250_sycl_kernel_lock_off.dat};
\addlegendentry{$W=1$}

\addplot[name path=up2, draw=none, forget plot] table [
  x index=0,
  y expr=\thisrowno{2}+\thisrowno{3},
  restrict expr to domain={\thisrowno{1}}{2:2}
] {pictures/data/bucket-mult/mi250-sycl/throughput_vs_bucket_mi250_sycl_kernel_lock_off.dat};

\addplot[name path=down2, draw=none, forget plot] table [
  x index=0,
  y expr=\thisrowno{2}-\thisrowno{3},
  restrict expr to domain={\thisrowno{1}}{2:2}
] {pictures/data/bucket-mult/mi250-sycl/throughput_vs_bucket_mi250_sycl_kernel_lock_off.dat};

\addplot[teal!80!black, fill opacity=0.18, forget plot] fill between[of=up2 and down2];

\addplot+[color=teal!80!black, solid, mark=square*] table [
  x index=0,
  y index=2,
  restrict expr to domain={\thisrowno{1}}{2:2}
] {pictures/data/bucket-mult/mi250-sycl/throughput_vs_bucket_mi250_sycl_kernel_lock_off.dat};
\addlegendentry{$W=2$}

\addplot[name path=up4, draw=none, forget plot] table [
  x index=0,
  y expr=\thisrowno{2}+\thisrowno{3},
  restrict expr to domain={\thisrowno{1}}{4:4}
] {pictures/data/bucket-mult/mi250-sycl/throughput_vs_bucket_mi250_sycl_kernel_lock_off.dat};

\addplot[name path=down4, draw=none, forget plot] table [
  x index=0,
  y expr=\thisrowno{2}-\thisrowno{3},
  restrict expr to domain={\thisrowno{1}}{4:4}
] {pictures/data/bucket-mult/mi250-sycl/throughput_vs_bucket_mi250_sycl_kernel_lock_off.dat};

\addplot[green!50!black, fill opacity=0.18, forget plot] fill between[of=up4 and down4];

\addplot+[color=green!50!black, solid, mark=triangle*] table [
  x index=0,
  y index=2,
  restrict expr to domain={\thisrowno{1}}{4:4}
] {pictures/data/bucket-mult/mi250-sycl/throughput_vs_bucket_mi250_sycl_kernel_lock_off.dat};
\addlegendentry{$W=4$}

\addplot[name path=up8, draw=none, forget plot] table [
  x index=0,
  y expr=\thisrowno{2}+\thisrowno{3},
  restrict expr to domain={\thisrowno{1}}{8:8}
] {pictures/data/bucket-mult/mi250-sycl/throughput_vs_bucket_mi250_sycl_kernel_lock_off.dat};

\addplot[name path=down8, draw=none, forget plot] table [
  x index=0,
  y expr=\thisrowno{2}-\thisrowno{3},
  restrict expr to domain={\thisrowno{1}}{8:8}
] {pictures/data/bucket-mult/mi250-sycl/throughput_vs_bucket_mi250_sycl_kernel_lock_off.dat};

\addplot[orange!90!black, fill opacity=0.18, forget plot] fill between[of=up8 and down8];

\addplot+[color=orange!90!black, solid, mark=diamond*] table [
  x index=0,
  y index=2,
  restrict expr to domain={\thisrowno{1}}{8:8}
] {pictures/data/bucket-mult/mi250-sycl/throughput_vs_bucket_mi250_sycl_kernel_lock_off.dat};
\addlegendentry{$W=8$}

\addplot[name path=up16, draw=none, forget plot] table [
  x index=0,
  y expr=\thisrowno{2}+\thisrowno{3},
  restrict expr to domain={\thisrowno{1}}{16:16}
] {pictures/data/bucket-mult/mi250-sycl/throughput_vs_bucket_mi250_sycl_kernel_lock_off.dat};

\addplot[name path=down16, draw=none, forget plot] table [
  x index=0,
  y expr=\thisrowno{2}-\thisrowno{3},
  restrict expr to domain={\thisrowno{1}}{16:16}
] {pictures/data/bucket-mult/mi250-sycl/throughput_vs_bucket_mi250_sycl_kernel_lock_off.dat};

\addplot[magenta!80!black, fill opacity=0.18, forget plot] fill between[of=up16 and down16];

\addplot+[color=magenta!80!black, solid, mark=pentagon*] table [
  x index=0,
  y index=2,
  restrict expr to domain={\thisrowno{1}}{16:16}
] {pictures/data/bucket-mult/mi250-sycl/throughput_vs_bucket_mi250_sycl_kernel_lock_off.dat};
\addlegendentry{$W=16$}

\nextgroupplot[
  ylabel={},
]

\addplot[name path=mbup1, draw=none] table [
  x index=0,
  y expr=\thisrowno{2}+\thisrowno{3},
  restrict expr to domain={\thisrowno{1}}{1:1}
] {pictures/data/bucket-mult/mi250-sycl/throughput_vs_bucket_mi250_sycl_kernel_lock_on.dat};

\addplot[name path=mbdown1, draw=none] table [
  x index=0,
  y expr=\thisrowno{2}-\thisrowno{3},
  restrict expr to domain={\thisrowno{1}}{1:1}
] {pictures/data/bucket-mult/mi250-sycl/throughput_vs_bucket_mi250_sycl_kernel_lock_on.dat};

\addplot[blue!80!black, fill opacity=0.18] fill between[of=mbup1 and mbdown1];

\addplot+[color=blue!80!black, solid, mark=*] table [
  x index=0,
  y index=2,
  restrict expr to domain={\thisrowno{1}}{1:1}
] {pictures/data/bucket-mult/mi250-sycl/throughput_vs_bucket_mi250_sycl_kernel_lock_on.dat};

\addplot[name path=mbup2, draw=none] table [
  x index=0,
  y expr=\thisrowno{2}+\thisrowno{3},
  restrict expr to domain={\thisrowno{1}}{2:2}
] {pictures/data/bucket-mult/mi250-sycl/throughput_vs_bucket_mi250_sycl_kernel_lock_on.dat};

\addplot[name path=mbdown2, draw=none] table [
  x index=0,
  y expr=\thisrowno{2}-\thisrowno{3},
  restrict expr to domain={\thisrowno{1}}{2:2}
] {pictures/data/bucket-mult/mi250-sycl/throughput_vs_bucket_mi250_sycl_kernel_lock_on.dat};

\addplot[teal!80!black, fill opacity=0.18] fill between[of=mbup2 and mbdown2];

\addplot+[color=teal!80!black, solid, mark=square*] table [
  x index=0,
  y index=2,
  restrict expr to domain={\thisrowno{1}}{2:2}
] {pictures/data/bucket-mult/mi250-sycl/throughput_vs_bucket_mi250_sycl_kernel_lock_on.dat};

\addplot[name path=mbup4, draw=none] table [
  x index=0,
  y expr=\thisrowno{2}+\thisrowno{3},
  restrict expr to domain={\thisrowno{1}}{4:4}
] {pictures/data/bucket-mult/mi250-sycl/throughput_vs_bucket_mi250_sycl_kernel_lock_on.dat};

\addplot[name path=mbdown4, draw=none] table [
  x index=0,
  y expr=\thisrowno{2}-\thisrowno{3},
  restrict expr to domain={\thisrowno{1}}{4:4}
] {pictures/data/bucket-mult/mi250-sycl/throughput_vs_bucket_mi250_sycl_kernel_lock_on.dat};

\addplot[green!50!black, fill opacity=0.18] fill between[of=mbup4 and mbdown4];

\addplot+[color=green!50!black, solid, mark=triangle*] table [
  x index=0,
  y index=2,
  restrict expr to domain={\thisrowno{1}}{4:4}
] {pictures/data/bucket-mult/mi250-sycl/throughput_vs_bucket_mi250_sycl_kernel_lock_on.dat};

\addplot[name path=mbup8, draw=none] table [
  x index=0,
  y expr=\thisrowno{2}+\thisrowno{3},
  restrict expr to domain={\thisrowno{1}}{8:8}
] {pictures/data/bucket-mult/mi250-sycl/throughput_vs_bucket_mi250_sycl_kernel_lock_on.dat};

\addplot[name path=mbdown8, draw=none] table [
  x index=0,
  y expr=\thisrowno{2}-\thisrowno{3},
  restrict expr to domain={\thisrowno{1}}{8:8}
] {pictures/data/bucket-mult/mi250-sycl/throughput_vs_bucket_mi250_sycl_kernel_lock_on.dat};

\addplot[orange!90!black, fill opacity=0.18] fill between[of=mbup8 and mbdown8];

\addplot+[color=orange!90!black, solid, mark=diamond*] table [
  x index=0,
  y index=2,
  restrict expr to domain={\thisrowno{1}}{8:8}
] {pictures/data/bucket-mult/mi250-sycl/throughput_vs_bucket_mi250_sycl_kernel_lock_on.dat};

\addplot[name path=mbup16, draw=none] table [
  x index=0,
  y expr=\thisrowno{2}+\thisrowno{3},
  restrict expr to domain={\thisrowno{1}}{16:16}
] {pictures/data/bucket-mult/mi250-sycl/throughput_vs_bucket_mi250_sycl_kernel_lock_on.dat};

\addplot[name path=mbdown16, draw=none] table [
  x index=0,
  y expr=\thisrowno{2}-\thisrowno{3},
  restrict expr to domain={\thisrowno{1}}{16:16}
] {pictures/data/bucket-mult/mi250-sycl/throughput_vs_bucket_mi250_sycl_kernel_lock_on.dat};

\addplot[magenta!80!black, fill opacity=0.18] fill between[of=mbup16 and mbdown16];

\addplot+[color=magenta!80!black, solid, mark=pentagon*] table [
  x index=0,
  y index=2,
  restrict expr to domain={\thisrowno{1}}{16:16}
] {pictures/data/bucket-mult/mi250-sycl/throughput_vs_bucket_mi250_sycl_kernel_lock_on.dat};

\end{groupplot}
\end{tikzpicture}

%% file: pictures/tikz/mem_work_gh200_cuda.tex
\begin{tikzpicture}
\begin{groupplot}[
  group style={
    group size=2 by 1,
    horizontal sep=1cm,
    ylabels at=edge left,
    xlabels at=edge bottom
  },
  width=0.44\textwidth,
  height=0.44\textwidth,
  xmin=1, xmax=16,
  xmode=log,
  log basis x=2,
  xtick={1,2,4,8,16},
  xticklabels={1,2,4,8,16},
  xticklabel style={/pgf/number format/fixed},
  ymin=0, ymax=7000,
  ytick={0,2000,4000,6000},
  xlabel={Memory Budget},
  x unit={\unit{\giga\byte}},
  ylabel={Throughput},
  y unit={ligands/\unit{\second}},
  grid=both,
  grid style={draw=gray!20},
  major grid style={draw=gray!30},
  unbounded coords=jump,
  tick label style={font=\footnotesize},
  label style={font=\footnotesize},
  title style={font=\footnotesize},
  legend style={font=\footnotesize, draw=none, fill=none, column sep=8pt},
  every axis plot/.append style={line width=1.1pt, mark size=2.2pt},
]

\nextgroupplot[
  title={Occupancy-guided},
  legend to name=memworklegendgh200cuda,
  legend columns=3,
]

\addplot[name path=up1, draw=none, forget plot] table [
  x index=0,
  y expr=\thisrowno{2}+\thisrowno{3},
  restrict expr to domain={\thisrowno{1}}{1:1}
] {pictures/data/mem-work/gh200-cuda/throughput_vs_mem_gh200_cuda_aligned_per_gpu.dat};

\addplot[name path=down1, draw=none, forget plot] table [
  x index=0,
  y expr=\thisrowno{2}-\thisrowno{3},
  restrict expr to domain={\thisrowno{1}}{1:1}
] {pictures/data/mem-work/gh200-cuda/throughput_vs_mem_gh200_cuda_aligned_per_gpu.dat};

\addplot[blue!80!black, fill opacity=0.18, forget plot] fill between[of=up1 and down1];

\addplot+[color=blue!80!black, solid, mark=*] table [
  x index=0,
  y index=2,
  restrict expr to domain={\thisrowno{1}}{1:1}
] {pictures/data/mem-work/gh200-cuda/throughput_vs_mem_gh200_cuda_aligned_per_gpu.dat};
\addlegendentry{$W=1$}

\addplot[name path=up2, draw=none, forget plot] table [
  x index=0,
  y expr=\thisrowno{2}+\thisrowno{3},
  restrict expr to domain={\thisrowno{1}}{2:2}
] {pictures/data/mem-work/gh200-cuda/throughput_vs_mem_gh200_cuda_aligned_per_gpu.dat};

\addplot[name path=down2, draw=none, forget plot] table [
  x index=0,
  y expr=\thisrowno{2}-\thisrowno{3},
  restrict expr to domain={\thisrowno{1}}{2:2}
] {pictures/data/mem-work/gh200-cuda/throughput_vs_mem_gh200_cuda_aligned_per_gpu.dat};

\addplot[teal!80!black, fill opacity=0.18, forget plot] fill between[of=up2 and down2];

\addplot+[color=teal!80!black, solid, mark=square*] table [
  x index=0,
  y index=2,
  restrict expr to domain={\thisrowno{1}}{2:2}
] {pictures/data/mem-work/gh200-cuda/throughput_vs_mem_gh200_cuda_aligned_per_gpu.dat};
\addlegendentry{$W=2$}

\addplot[name path=up4, draw=none, forget plot] table [
  x index=0,
  y expr=\thisrowno{2}+\thisrowno{3},
  restrict expr to domain={\thisrowno{1}}{4:4}
] {pictures/data/mem-work/gh200-cuda/throughput_vs_mem_gh200_cuda_aligned_per_gpu.dat};

\addplot[name path=down4, draw=none, forget plot] table [
  x index=0,
  y expr=\thisrowno{2}-\thisrowno{3},
  restrict expr to domain={\thisrowno{1}}{4:4}
] {pictures/data/mem-work/gh200-cuda/throughput_vs_mem_gh200_cuda_aligned_per_gpu.dat};

\addplot[green!50!black, fill opacity=0.18, forget plot] fill between[of=up4 and down4];

\addplot+[color=green!50!black, solid, mark=triangle*] table [
  x index=0,
  y index=2,
  restrict expr to domain={\thisrowno{1}}{4:4}
] {pictures/data/mem-work/gh200-cuda/throughput_vs_mem_gh200_cuda_aligned_per_gpu.dat};
\addlegendentry{$W=4$}

\addplot[name path=up8, draw=none, forget plot] table [
  x index=0,
  y expr=\thisrowno{2}+\thisrowno{3},
  restrict expr to domain={\thisrowno{1}}{8:8}
] {pictures/data/mem-work/gh200-cuda/throughput_vs_mem_gh200_cuda_aligned_per_gpu.dat};

\addplot[name path=down8, draw=none, forget plot] table [
  x index=0,
  y expr=\thisrowno{2}-\thisrowno{3},
  restrict expr to domain={\thisrowno{1}}{8:8}
] {pictures/data/mem-work/gh200-cuda/throughput_vs_mem_gh200_cuda_aligned_per_gpu.dat};

\addplot[orange!90!black, fill opacity=0.18, forget plot] fill between[of=up8 and down8];

\addplot+[color=orange!90!black, solid, mark=diamond*] table [
  x index=0,
  y index=2,
  restrict expr to domain={\thisrowno{1}}{8:8}
] {pictures/data/mem-work/gh200-cuda/throughput_vs_mem_gh200_cuda_aligned_per_gpu.dat};
\addlegendentry{$W=8$}

\addplot[name path=up16, draw=none, forget plot] table [
  x index=0,
  y expr=\thisrowno{2}+\thisrowno{3},
  restrict expr to domain={\thisrowno{1}}{16:16}
] {pictures/data/mem-work/gh200-cuda/throughput_vs_mem_gh200_cuda_aligned_per_gpu.dat};

\addplot[name path=down16, draw=none, forget plot] table [
  x index=0,
  y expr=\thisrowno{2}-\thisrowno{3},
  restrict expr to domain={\thisrowno{1}}{16:16}
] {pictures/data/mem-work/gh200-cuda/throughput_vs_mem_gh200_cuda_aligned_per_gpu.dat};

\addplot[magenta!80!black, fill opacity=0.18, forget plot] fill between[of=up16 and down16];

\addplot+[color=magenta!80!black, solid, mark=pentagon*] table [
  x index=0,
  y index=2,
  restrict expr to domain={\thisrowno{1}}{16:16}
] {pictures/data/mem-work/gh200-cuda/throughput_vs_mem_gh200_cuda_aligned_per_gpu.dat};
\addlegendentry{$W=16$}

\nextgroupplot[
  title={Memory-bounded},
  ylabel={},
]

\addplot[name path=mbup1, draw=none] table [
  x index=0,
  y expr=\thisrowno{2}+\thisrowno{3},
  restrict expr to domain={\thisrowno{1}}{1:1}
] {pictures/data/mem-work/gh200-cuda/throughput_vs_mem_gh200_cuda_memory_only.dat};

\addplot[name path=mbdown1, draw=none] table [
  x index=0,
  y expr=\thisrowno{2}-\thisrowno{3},
  restrict expr to domain={\thisrowno{1}}{1:1}
] {pictures/data/mem-work/gh200-cuda/throughput_vs_mem_gh200_cuda_memory_only.dat};

\addplot[blue!80!black, fill opacity=0.18] fill between[of=mbup1 and mbdown1];

\addplot+[color=blue!80!black, solid, mark=*] table [
  x index=0,
  y index=2,
  restrict expr to domain={\thisrowno{1}}{1:1}
] {pictures/data/mem-work/gh200-cuda/throughput_vs_mem_gh200_cuda_memory_only.dat};

\addplot[name path=mbup2, draw=none] table [
  x index=0,
  y expr=\thisrowno{2}+\thisrowno{3},
  restrict expr to domain={\thisrowno{1}}{2:2}
] {pictures/data/mem-work/gh200-cuda/throughput_vs_mem_gh200_cuda_memory_only.dat};

\addplot[name path=mbdown2, draw=none] table [
  x index=0,
  y expr=\thisrowno{2}-\thisrowno{3},
  restrict expr to domain={\thisrowno{1}}{2:2}
] {pictures/data/mem-work/gh200-cuda/throughput_vs_mem_gh200_cuda_memory_only.dat};

\addplot[teal!80!black, fill opacity=0.18] fill between[of=mbup2 and mbdown2];

\addplot+[color=teal!80!black, solid, mark=square*] table [
  x index=0,
  y index=2,
  restrict expr to domain={\thisrowno{1}}{2:2}
] {pictures/data/mem-work/gh200-cuda/throughput_vs_mem_gh200_cuda_memory_only.dat};

\addplot[name path=mbup4, draw=none] table [
  x index=0,
  y expr=\thisrowno{2}+\thisrowno{3},
  restrict expr to domain={\thisrowno{1}}{4:4}
] {pictures/data/mem-work/gh200-cuda/throughput_vs_mem_gh200_cuda_memory_only.dat};

\addplot[name path=mbdown4, draw=none] table [
  x index=0,
  y expr=\thisrowno{2}-\thisrowno{3},
  restrict expr to domain={\thisrowno{1}}{4:4}
] {pictures/data/mem-work/gh200-cuda/throughput_vs_mem_gh200_cuda_memory_only.dat};

\addplot[green!50!black, fill opacity=0.18] fill between[of=mbup4 and mbdown4];

\addplot+[color=green!50!black, solid, mark=triangle*] table [
  x index=0,
  y index=2,
  restrict expr to domain={\thisrowno{1}}{4:4}
] {pictures/data/mem-work/gh200-cuda/throughput_vs_mem_gh200_cuda_memory_only.dat};

\addplot[name path=mbup8, draw=none] table [
  x index=0,
  y expr=\thisrowno{2}+\thisrowno{3},
  restrict expr to domain={\thisrowno{1}}{8:8}
] {pictures/data/mem-work/gh200-cuda/throughput_vs_mem_gh200_cuda_memory_only.dat};

\addplot[name path=mbdown8, draw=none] table [
  x index=0,
  y expr=\thisrowno{2}-\thisrowno{3},
  restrict expr to domain={\thisrowno{1}}{8:8}
] {pictures/data/mem-work/gh200-cuda/throughput_vs_mem_gh200_cuda_memory_only.dat};

\addplot[orange!90!black, fill opacity=0.18] fill between[of=mbup8 and mbdown8];

\addplot+[color=orange!90!black, solid, mark=diamond*] table [
  x index=0,
  y index=2,
  restrict expr to domain={\thisrowno{1}}{8:8}
] {pictures/data/mem-work/gh200-cuda/throughput_vs_mem_gh200_cuda_memory_only.dat};

\addplot[name path=mbup16, draw=none] table [
  x index=0,
  y expr=\thisrowno{2}+\thisrowno{3},
  restrict expr to domain={\thisrowno{1}}{16:16}
] {pictures/data/mem-work/gh200-cuda/throughput_vs_mem_gh200_cuda_memory_only.dat};

\addplot[name path=mbdown16, draw=none] table [
  x index=0,
  y expr=\thisrowno{2}-\thisrowno{3},
  restrict expr to domain={\thisrowno{1}}{16:16}
] {pictures/data/mem-work/gh200-cuda/throughput_vs_mem_gh200_cuda_memory_only.dat};

\addplot[magenta!80!black, fill opacity=0.18] fill between[of=mbup16 and mbdown16];

\addplot+[color=magenta!80!black, solid, mark=pentagon*] table [
  x index=0,
  y index=2,
  restrict expr to domain={\thisrowno{1}}{16:16}
] {pictures/data/mem-work/gh200-cuda/throughput_vs_mem_gh200_cuda_memory_only.dat};

\end{groupplot}
\end{tikzpicture}

%% file: pictures/tikz/mem_work_gh200_hip.tex
\begin{tikzpicture}
\begin{groupplot}[
  group style={
    group size=2 by 1,
    horizontal sep=1cm,
    ylabels at=edge left,
    xlabels at=edge bottom
  },
  width=0.44\textwidth,
  height=0.44\textwidth,
  xmin=1, xmax=16,
  xmode=log,
  log basis x=2,
  xtick={1,2,4,8,16},
  xticklabels={1,2,4,8,16},
  xticklabel style={/pgf/number format/fixed},
  ymin=0, ymax=7000,
  ytick={0,2000,4000,6000},
  xlabel={Memory Budget},
  x unit={\unit{\giga\byte}},
  ylabel={Throughput},
  y unit={ligands/\unit{\second}},
  grid=both,
  grid style={draw=gray!20},
  major grid style={draw=gray!30},
  unbounded coords=jump,
  tick label style={font=\footnotesize},
  label style={font=\footnotesize},
  title style={font=\footnotesize},
  legend style={font=\footnotesize, draw=none, fill=none, column sep=8pt},
  every axis plot/.append style={line width=1.1pt, mark size=2.2pt},
]

\nextgroupplot[
  title={Occupancy-guided},
  legend to name=memworklegendgh200hip,
  legend columns=3,
]

\addplot[name path=up1, draw=none, forget plot] table [
  x index=0,
  y expr=\thisrowno{2}+\thisrowno{3},
  restrict expr to domain={\thisrowno{1}}{1:1}
] {pictures/data/mem-work/gh200-hip/throughput_vs_mem_gh200_hip_aligned_per_gpu.dat};

\addplot[name path=down1, draw=none, forget plot] table [
  x index=0,
  y expr=\thisrowno{2}-\thisrowno{3},
  restrict expr to domain={\thisrowno{1}}{1:1}
] {pictures/data/mem-work/gh200-hip/throughput_vs_mem_gh200_hip_aligned_per_gpu.dat};

\addplot[blue!80!black, fill opacity=0.18, forget plot] fill between[of=up1 and down1];

\addplot+[color=blue!80!black, solid, mark=*] table [
  x index=0,
  y index=2,
  restrict expr to domain={\thisrowno{1}}{1:1}
] {pictures/data/mem-work/gh200-hip/throughput_vs_mem_gh200_hip_aligned_per_gpu.dat};
\addlegendentry{$W=1$}

\addplot[name path=up2, draw=none, forget plot] table [
  x index=0,
  y expr=\thisrowno{2}+\thisrowno{3},
  restrict expr to domain={\thisrowno{1}}{2:2}
] {pictures/data/mem-work/gh200-hip/throughput_vs_mem_gh200_hip_aligned_per_gpu.dat};

\addplot[name path=down2, draw=none, forget plot] table [
  x index=0,
  y expr=\thisrowno{2}-\thisrowno{3},
  restrict expr to domain={\thisrowno{1}}{2:2}
] {pictures/data/mem-work/gh200-hip/throughput_vs_mem_gh200_hip_aligned_per_gpu.dat};

\addplot[teal!80!black, fill opacity=0.18, forget plot] fill between[of=up2 and down2];

\addplot+[color=teal!80!black, solid, mark=square*] table [
  x index=0,
  y index=2,
  restrict expr to domain={\thisrowno{1}}{2:2}
] {pictures/data/mem-work/gh200-hip/throughput_vs_mem_gh200_hip_aligned_per_gpu.dat};
\addlegendentry{$W=2$}

\addplot[name path=up4, draw=none, forget plot] table [
  x index=0,
  y expr=\thisrowno{2}+\thisrowno{3},
  restrict expr to domain={\thisrowno{1}}{4:4}
] {pictures/data/mem-work/gh200-hip/throughput_vs_mem_gh200_hip_aligned_per_gpu.dat};

\addplot[name path=down4, draw=none, forget plot] table [
  x index=0,
  y expr=\thisrowno{2}-\thisrowno{3},
  restrict expr to domain={\thisrowno{1}}{4:4}
] {pictures/data/mem-work/gh200-hip/throughput_vs_mem_gh200_hip_aligned_per_gpu.dat};

\addplot[green!50!black, fill opacity=0.18, forget plot] fill between[of=up4 and down4];

\addplot+[color=green!50!black, solid, mark=triangle*] table [
  x index=0,
  y index=2,
  restrict expr to domain={\thisrowno{1}}{4:4}
] {pictures/data/mem-work/gh200-hip/throughput_vs_mem_gh200_hip_aligned_per_gpu.dat};
\addlegendentry{$W=4$}

\addplot[name path=up8, draw=none, forget plot] table [
  x index=0,
  y expr=\thisrowno{2}+\thisrowno{3},
  restrict expr to domain={\thisrowno{1}}{8:8}
] {pictures/data/mem-work/gh200-hip/throughput_vs_mem_gh200_hip_aligned_per_gpu.dat};

\addplot[name path=down8, draw=none, forget plot] table [
  x index=0,
  y expr=\thisrowno{2}-\thisrowno{3},
  restrict expr to domain={\thisrowno{1}}{8:8}
] {pictures/data/mem-work/gh200-hip/throughput_vs_mem_gh200_hip_aligned_per_gpu.dat};

\addplot[orange!90!black, fill opacity=0.18, forget plot] fill between[of=up8 and down8];

\addplot+[color=orange!90!black, solid, mark=diamond*] table [
  x index=0,
  y index=2,
  restrict expr to domain={\thisrowno{1}}{8:8}
] {pictures/data/mem-work/gh200-hip/throughput_vs_mem_gh200_hip_aligned_per_gpu.dat};
\addlegendentry{$W=8$}

\addplot[name path=up16, draw=none, forget plot] table [
  x index=0,
  y expr=\thisrowno{2}+\thisrowno{3},
  restrict expr to domain={\thisrowno{1}}{16:16}
] {pictures/data/mem-work/gh200-hip/throughput_vs_mem_gh200_hip_aligned_per_gpu.dat};

\addplot[name path=down16, draw=none, forget plot] table [
  x index=0,
  y expr=\thisrowno{2}-\thisrowno{3},
  restrict expr to domain={\thisrowno{1}}{16:16}
] {pictures/data/mem-work/gh200-hip/throughput_vs_mem_gh200_hip_aligned_per_gpu.dat};

\addplot[magenta!80!black, fill opacity=0.18, forget plot] fill between[of=up16 and down16];

\addplot+[color=magenta!80!black, solid, mark=pentagon*] table [
  x index=0,
  y index=2,
  restrict expr to domain={\thisrowno{1}}{16:16}
] {pictures/data/mem-work/gh200-hip/throughput_vs_mem_gh200_hip_aligned_per_gpu.dat};
\addlegendentry{$W=16$}

\nextgroupplot[
  title={Memory-bounded},
  ylabel={},
]

\addplot[name path=mbup1, draw=none] table [
  x index=0,
  y expr=\thisrowno{2}+\thisrowno{3},
  restrict expr to domain={\thisrowno{1}}{1:1}
] {pictures/data/mem-work/gh200-hip/throughput_vs_mem_gh200_hip_memory_only.dat};

\addplot[name path=mbdown1, draw=none] table [
  x index=0,
  y expr=\thisrowno{2}-\thisrowno{3},
  restrict expr to domain={\thisrowno{1}}{1:1}
] {pictures/data/mem-work/gh200-hip/throughput_vs_mem_gh200_hip_memory_only.dat};

\addplot[blue!80!black, fill opacity=0.18] fill between[of=mbup1 and mbdown1];

\addplot+[color=blue!80!black, solid, mark=*] table [
  x index=0,
  y index=2,
  restrict expr to domain={\thisrowno{1}}{1:1}
] {pictures/data/mem-work/gh200-hip/throughput_vs_mem_gh200_hip_memory_only.dat};

\addplot[name path=mbup2, draw=none] table [
  x index=0,
  y expr=\thisrowno{2}+\thisrowno{3},
  restrict expr to domain={\thisrowno{1}}{2:2}
] {pictures/data/mem-work/gh200-hip/throughput_vs_mem_gh200_hip_memory_only.dat};

\addplot[name path=mbdown2, draw=none] table [
  x index=0,
  y expr=\thisrowno{2}-\thisrowno{3},
  restrict expr to domain={\thisrowno{1}}{2:2}
] {pictures/data/mem-work/gh200-hip/throughput_vs_mem_gh200_hip_memory_only.dat};

\addplot[teal!80!black, fill opacity=0.18] fill between[of=mbup2 and mbdown2];

\addplot+[color=teal!80!black, solid, mark=square*] table [
  x index=0,
  y index=2,
  restrict expr to domain={\thisrowno{1}}{2:2}
] {pictures/data/mem-work/gh200-hip/throughput_vs_mem_gh200_hip_memory_only.dat};

\addplot[name path=mbup4, draw=none] table [
  x index=0,
  y expr=\thisrowno{2}+\thisrowno{3},
  restrict expr to domain={\thisrowno{1}}{4:4}
] {pictures/data/mem-work/gh200-hip/throughput_vs_mem_gh200_hip_memory_only.dat};

\addplot[name path=mbdown4, draw=none] table [
  x index=0,
  y expr=\thisrowno{2}-\thisrowno{3},
  restrict expr to domain={\thisrowno{1}}{4:4}
] {pictures/data/mem-work/gh200-hip/throughput_vs_mem_gh200_hip_memory_only.dat};

\addplot[green!50!black, fill opacity=0.18] fill between[of=mbup4 and mbdown4];

\addplot+[color=green!50!black, solid, mark=triangle*] table [
  x index=0,
  y index=2,
  restrict expr to domain={\thisrowno{1}}{4:4}
] {pictures/data/mem-work/gh200-hip/throughput_vs_mem_gh200_hip_memory_only.dat};

\addplot[name path=mbup8, draw=none] table [
  x index=0,
  y expr=\thisrowno{2}+\thisrowno{3},
  restrict expr to domain={\thisrowno{1}}{8:8}
] {pictures/data/mem-work/gh200-hip/throughput_vs_mem_gh200_hip_memory_only.dat};

\addplot[name path=mbdown8, draw=none] table [
  x index=0,
  y expr=\thisrowno{2}-\thisrowno{3},
  restrict expr to domain={\thisrowno{1}}{8:8}
] {pictures/data/mem-work/gh200-hip/throughput_vs_mem_gh200_hip_memory_only.dat};

\addplot[orange!90!black, fill opacity=0.18] fill between[of=mbup8 and mbdown8];

\addplot+[color=orange!90!black, solid, mark=diamond*] table [
  x index=0,
  y index=2,
  restrict expr to domain={\thisrowno{1}}{8:8}
] {pictures/data/mem-work/gh200-hip/throughput_vs_mem_gh200_hip_memory_only.dat};

\addplot[name path=mbup16, draw=none] table [
  x index=0,
  y expr=\thisrowno{2}+\thisrowno{3},
  restrict expr to domain={\thisrowno{1}}{16:16}
] {pictures/data/mem-work/gh200-hip/throughput_vs_mem_gh200_hip_memory_only.dat};

\addplot[name path=mbdown16, draw=none] table [
  x index=0,
  y expr=\thisrowno{2}-\thisrowno{3},
  restrict expr to domain={\thisrowno{1}}{16:16}
] {pictures/data/mem-work/gh200-hip/throughput_vs_mem_gh200_hip_memory_only.dat};

\addplot[magenta!80!black, fill opacity=0.18] fill between[of=mbup16 and mbdown16];

\addplot+[color=magenta!80!black, solid, mark=pentagon*] table [
  x index=0,
  y index=2,
  restrict expr to domain={\thisrowno{1}}{16:16}
] {pictures/data/mem-work/gh200-hip/throughput_vs_mem_gh200_hip_memory_only.dat};

\end{groupplot}
\end{tikzpicture}

%% file: pictures/tikz/mem_work_gh200_sycl.tex
\begin{tikzpicture}
\begin{groupplot}[
  group style={
    group size=2 by 1,
    horizontal sep=1cm,
    ylabels at=edge left,
    xlabels at=edge bottom
  },
  width=0.44\textwidth,
  height=0.44\textwidth,
  xmin=1, xmax=16,
  xmode=log,
  log basis x=2,
  xtick={1,2,4,8,16},
  xticklabels={1,2,4,8,16},
  xticklabel style={/pgf/number format/fixed},
  ymin=0, ymax=3800,
  ytick={0,1000,2000,3000},
  xlabel={Memory Budget},
  x unit={\unit{\giga\byte}},
  ylabel={Throughput},
  y unit={ligands/\unit{\second}},
  grid=both,
  grid style={draw=gray!20},
  major grid style={draw=gray!30},
  unbounded coords=jump,
  tick label style={font=\footnotesize},
  label style={font=\footnotesize},
  title style={font=\footnotesize},
  legend style={font=\footnotesize, draw=none, fill=none, column sep=8pt},
  every axis plot/.append style={line width=1.1pt, mark size=2.2pt},
]

\nextgroupplot[
  legend to name=memworklegendgh200sycl,
  legend columns=3,
]

\addplot[name path=up1, draw=none, forget plot] table [
  x index=0,
  y expr=\thisrowno{2}+\thisrowno{3},
  restrict expr to domain={\thisrowno{1}}{1:1}
] {pictures/data/mem-work/gh200-sycl/throughput_vs_mem_gh200_sycl_aligned_per_gpu.dat};

\addplot[name path=down1, draw=none, forget plot] table [
  x index=0,
  y expr=\thisrowno{2}-\thisrowno{3},
  restrict expr to domain={\thisrowno{1}}{1:1}
] {pictures/data/mem-work/gh200-sycl/throughput_vs_mem_gh200_sycl_aligned_per_gpu.dat};

\addplot[blue!80!black, fill opacity=0.18, forget plot] fill between[of=up1 and down1];

\addplot+[color=blue!80!black, solid, mark=*] table [
  x index=0,
  y index=2,
  restrict expr to domain={\thisrowno{1}}{1:1}
] {pictures/data/mem-work/gh200-sycl/throughput_vs_mem_gh200_sycl_aligned_per_gpu.dat};
\addlegendentry{$W=1$}

\addplot[name path=up2, draw=none, forget plot] table [
  x index=0,
  y expr=\thisrowno{2}+\thisrowno{3},
  restrict expr to domain={\thisrowno{1}}{2:2}
] {pictures/data/mem-work/gh200-sycl/throughput_vs_mem_gh200_sycl_aligned_per_gpu.dat};

\addplot[name path=down2, draw=none, forget plot] table [
  x index=0,
  y expr=\thisrowno{2}-\thisrowno{3},
  restrict expr to domain={\thisrowno{1}}{2:2}
] {pictures/data/mem-work/gh200-sycl/throughput_vs_mem_gh200_sycl_aligned_per_gpu.dat};

\addplot[teal!80!black, fill opacity=0.18, forget plot] fill between[of=up2 and down2];

\addplot+[color=teal!80!black, solid, mark=square*] table [
  x index=0,
  y index=2,
  restrict expr to domain={\thisrowno{1}}{2:2}
] {pictures/data/mem-work/gh200-sycl/throughput_vs_mem_gh200_sycl_aligned_per_gpu.dat};
\addlegendentry{$W=2$}

\addplot[name path=up4, draw=none, forget plot] table [
  x index=0,
  y expr=\thisrowno{2}+\thisrowno{3},
  restrict expr to domain={\thisrowno{1}}{4:4}
] {pictures/data/mem-work/gh200-sycl/throughput_vs_mem_gh200_sycl_aligned_per_gpu.dat};

\addplot[name path=down4, draw=none, forget plot] table [
  x index=0,
  y expr=\thisrowno{2}-\thisrowno{3},
  restrict expr to domain={\thisrowno{1}}{4:4}
] {pictures/data/mem-work/gh200-sycl/throughput_vs_mem_gh200_sycl_aligned_per_gpu.dat};

\addplot[green!50!black, fill opacity=0.18, forget plot] fill between[of=up4 and down4];

\addplot+[color=green!50!black, solid, mark=triangle*] table [
  x index=0,
  y index=2,
  restrict expr to domain={\thisrowno{1}}{4:4}
] {pictures/data/mem-work/gh200-sycl/throughput_vs_mem_gh200_sycl_aligned_per_gpu.dat};
\addlegendentry{$W=4$}

\addplot[name path=up8, draw=none, forget plot] table [
  x index=0,
  y expr=\thisrowno{2}+\thisrowno{3},
  restrict expr to domain={\thisrowno{1}}{8:8}
] {pictures/data/mem-work/gh200-sycl/throughput_vs_mem_gh200_sycl_aligned_per_gpu.dat};

\addplot[name path=down8, draw=none, forget plot] table [
  x index=0,
  y expr=\thisrowno{2}-\thisrowno{3},
  restrict expr to domain={\thisrowno{1}}{8:8}
] {pictures/data/mem-work/gh200-sycl/throughput_vs_mem_gh200_sycl_aligned_per_gpu.dat};

\addplot[orange!90!black, fill opacity=0.18, forget plot] fill between[of=up8 and down8];

\addplot+[color=orange!90!black, solid, mark=diamond*] table [
  x index=0,
  y index=2,
  restrict expr to domain={\thisrowno{1}}{8:8}
] {pictures/data/mem-work/gh200-sycl/throughput_vs_mem_gh200_sycl_aligned_per_gpu.dat};
\addlegendentry{$W=8$}

\addplot[name path=up16, draw=none, forget plot] table [
  x index=0,
  y expr=\thisrowno{2}+\thisrowno{3},
  restrict expr to domain={\thisrowno{1}}{16:16}
] {pictures/data/mem-work/gh200-sycl/throughput_vs_mem_gh200_sycl_aligned_per_gpu.dat};

\addplot[name path=down16, draw=none, forget plot] table [
  x index=0,
  y expr=\thisrowno{2}-\thisrowno{3},
  restrict expr to domain={\thisrowno{1}}{16:16}
] {pictures/data/mem-work/gh200-sycl/throughput_vs_mem_gh200_sycl_aligned_per_gpu.dat};

\addplot[magenta!80!black, fill opacity=0.18, forget plot] fill between[of=up16 and down16];

\addplot+[color=magenta!80!black, solid, mark=pentagon*] table [
  x index=0,
  y index=2,
  restrict expr to domain={\thisrowno{1}}{16:16}
] {pictures/data/mem-work/gh200-sycl/throughput_vs_mem_gh200_sycl_aligned_per_gpu.dat};
\addlegendentry{$W=16$}

\nextgroupplot[
  ylabel={},
]

\addplot[name path=mbup1, draw=none] table [
  x index=0,
  y expr=\thisrowno{2}+\thisrowno{3},
  restrict expr to domain={\thisrowno{1}}{1:1}
] {pictures/data/mem-work/gh200-sycl/throughput_vs_mem_gh200_sycl_memory_only.dat};

\addplot[name path=mbdown1, draw=none] table [
  x index=0,
  y expr=\thisrowno{2}-\thisrowno{3},
  restrict expr to domain={\thisrowno{1}}{1:1}
] {pictures/data/mem-work/gh200-sycl/throughput_vs_mem_gh200_sycl_memory_only.dat};

\addplot[blue!80!black, fill opacity=0.18] fill between[of=mbup1 and mbdown1];

\addplot+[color=blue!80!black, solid, mark=*] table [
  x index=0,
  y index=2,
  restrict expr to domain={\thisrowno{1}}{1:1}
] {pictures/data/mem-work/gh200-sycl/throughput_vs_mem_gh200_sycl_memory_only.dat};

\addplot[name path=mbup2, draw=none] table [
  x index=0,
  y expr=\thisrowno{2}+\thisrowno{3},
  restrict expr to domain={\thisrowno{1}}{2:2}
] {pictures/data/mem-work/gh200-sycl/throughput_vs_mem_gh200_sycl_memory_only.dat};

\addplot[name path=mbdown2, draw=none] table [
  x index=0,
  y expr=\thisrowno{2}-\thisrowno{3},
  restrict expr to domain={\thisrowno{1}}{2:2}
] {pictures/data/mem-work/gh200-sycl/throughput_vs_mem_gh200_sycl_memory_only.dat};

\addplot[teal!80!black, fill opacity=0.18] fill between[of=mbup2 and mbdown2];

\addplot+[color=teal!80!black, solid, mark=square*] table [
  x index=0,
  y index=2,
  restrict expr to domain={\thisrowno{1}}{2:2}
] {pictures/data/mem-work/gh200-sycl/throughput_vs_mem_gh200_sycl_memory_only.dat};

\addplot[name path=mbup4, draw=none] table [
  x index=0,
  y expr=\thisrowno{2}+\thisrowno{3},
  restrict expr to domain={\thisrowno{1}}{4:4}
] {pictures/data/mem-work/gh200-sycl/throughput_vs_mem_gh200_sycl_memory_only.dat};

\addplot[name path=mbdown4, draw=none] table [
  x index=0,
  y expr=\thisrowno{2}-\thisrowno{3},
  restrict expr to domain={\thisrowno{1}}{4:4}
] {pictures/data/mem-work/gh200-sycl/throughput_vs_mem_gh200_sycl_memory_only.dat};

\addplot[green!50!black, fill opacity=0.18] fill between[of=mbup4 and mbdown4];

\addplot+[color=green!50!black, solid, mark=triangle*] table [
  x index=0,
  y index=2,
  restrict expr to domain={\thisrowno{1}}{4:4}
] {pictures/data/mem-work/gh200-sycl/throughput_vs_mem_gh200_sycl_memory_only.dat};

\addplot[name path=mbup8, draw=none] table [
  x index=0,
  y expr=\thisrowno{2}+\thisrowno{3},
  restrict expr to domain={\thisrowno{1}}{8:8}
] {pictures/data/mem-work/gh200-sycl/throughput_vs_mem_gh200_sycl_memory_only.dat};

\addplot[name path=mbdown8, draw=none] table [
  x index=0,
  y expr=\thisrowno{2}-\thisrowno{3},
  restrict expr to domain={\thisrowno{1}}{8:8}
] {pictures/data/mem-work/gh200-sycl/throughput_vs_mem_gh200_sycl_memory_only.dat};

\addplot[orange!90!black, fill opacity=0.18] fill between[of=mbup8 and mbdown8];

\addplot+[color=orange!90!black, solid, mark=diamond*] table [
  x index=0,
  y index=2,
  restrict expr to domain={\thisrowno{1}}{8:8}
] {pictures/data/mem-work/gh200-sycl/throughput_vs_mem_gh200_sycl_memory_only.dat};

\addplot[name path=mbup16, draw=none] table [
  x index=0,
  y expr=\thisrowno{2}+\thisrowno{3},
  restrict expr to domain={\thisrowno{1}}{16:16}
] {pictures/data/mem-work/gh200-sycl/throughput_vs_mem_gh200_sycl_memory_only.dat};

\addplot[name path=mbdown16, draw=none] table [
  x index=0,
  y expr=\thisrowno{2}-\thisrowno{3},
  restrict expr to domain={\thisrowno{1}}{16:16}
] {pictures/data/mem-work/gh200-sycl/throughput_vs_mem_gh200_sycl_memory_only.dat};

\addplot[magenta!80!black, fill opacity=0.18] fill between[of=mbup16 and mbdown16];

\addplot+[color=magenta!80!black, solid, mark=pentagon*] table [
  x index=0,
  y index=2,
  restrict expr to domain={\thisrowno{1}}{16:16}
] {pictures/data/mem-work/gh200-sycl/throughput_vs_mem_gh200_sycl_memory_only.dat};

\end{groupplot}
\end{tikzpicture}

%% file: pictures/tikz/mem_work_mi250_hip.tex
\begin{tikzpicture}
\begin{groupplot}[
  group style={
    group size=2 by 1,
    horizontal sep=1cm,
    ylabels at=edge left,
    xlabels at=edge bottom
  },
  width=0.44\textwidth,
  height=0.44\textwidth,
  xmin=1, xmax=16,
  xmode=log,
  log basis x=2,
  xtick={1,2,4,8,16},
  xticklabels={1,2,4,8,16},
  xticklabel style={/pgf/number format/fixed},
  ymin=0, ymax=1500,
  ytick={0,500,1000,1500},
  xlabel={Memory Budget},
  x unit={\unit{\giga\byte}},
  ylabel={Throughput},
  y unit={ligands/\unit{\second}},
  grid=both,
  grid style={draw=gray!20},
  major grid style={draw=gray!30},
  unbounded coords=jump,
  tick label style={font=\footnotesize},
  label style={font=\footnotesize},
  title style={font=\footnotesize},
  legend style={font=\footnotesize, draw=none, fill=none, column sep=8pt},
  every axis plot/.append style={line width=1.1pt, mark size=2.2pt},
]

\nextgroupplot[
  legend to name=memworklegendgh200sycl,
  legend columns=3,
]

\addplot[name path=up1, draw=none, forget plot] table [
  x index=0,
  y expr=\thisrowno{2}+\thisrowno{3},
  restrict expr to domain={\thisrowno{1}}{1:1}
] {pictures/data/mem-work/mi250-hip/throughput_vs_mem_mi250_hip_aligned_per_gpu.dat};

\addplot[name path=down1, draw=none, forget plot] table [
  x index=0,
  y expr=\thisrowno{2}-\thisrowno{3},
  restrict expr to domain={\thisrowno{1}}{1:1}
] {pictures/data/mem-work/mi250-hip/throughput_vs_mem_mi250_hip_aligned_per_gpu.dat};

\addplot[blue!80!black, fill opacity=0.18, forget plot] fill between[of=up1 and down1];

\addplot+[color=blue!80!black, solid, mark=*] table [
  x index=0,
  y index=2,
  restrict expr to domain={\thisrowno{1}}{1:1}
] {pictures/data/mem-work/mi250-hip/throughput_vs_mem_mi250_hip_aligned_per_gpu.dat};
\addlegendentry{$W=1$}

\addplot[name path=up2, draw=none, forget plot] table [
  x index=0,
  y expr=\thisrowno{2}+\thisrowno{3},
  restrict expr to domain={\thisrowno{1}}{2:2}
] {pictures/data/mem-work/mi250-hip/throughput_vs_mem_mi250_hip_aligned_per_gpu.dat};

\addplot[name path=down2, draw=none, forget plot] table [
  x index=0,
  y expr=\thisrowno{2}-\thisrowno{3},
  restrict expr to domain={\thisrowno{1}}{2:2}
] {pictures/data/mem-work/mi250-hip/throughput_vs_mem_mi250_hip_aligned_per_gpu.dat};

\addplot[teal!80!black, fill opacity=0.18, forget plot] fill between[of=up2 and down2];

\addplot+[color=teal!80!black, solid, mark=square*] table [
  x index=0,
  y index=2,
  restrict expr to domain={\thisrowno{1}}{2:2}
] {pictures/data/mem-work/mi250-hip/throughput_vs_mem_mi250_hip_aligned_per_gpu.dat};
\addlegendentry{$W=2$}

\addplot[name path=up4, draw=none, forget plot] table [
  x index=0,
  y expr=\thisrowno{2}+\thisrowno{3},
  restrict expr to domain={\thisrowno{1}}{4:4}
] {pictures/data/mem-work/mi250-hip/throughput_vs_mem_mi250_hip_aligned_per_gpu.dat};

\addplot[name path=down4, draw=none, forget plot] table [
  x index=0,
  y expr=\thisrowno{2}-\thisrowno{3},
  restrict expr to domain={\thisrowno{1}}{4:4}
] {pictures/data/mem-work/mi250-hip/throughput_vs_mem_mi250_hip_aligned_per_gpu.dat};

\addplot[green!50!black, fill opacity=0.18, forget plot] fill between[of=up4 and down4];

\addplot+[color=green!50!black, solid, mark=triangle*] table [
  x index=0,
  y index=2,
  restrict expr to domain={\thisrowno{1}}{4:4}
] {pictures/data/mem-work/mi250-hip/throughput_vs_mem_mi250_hip_aligned_per_gpu.dat};
\addlegendentry{$W=4$}

\addplot[name path=up8, draw=none, forget plot] table [
  x index=0,
  y expr=\thisrowno{2}+\thisrowno{3},
  restrict expr to domain={\thisrowno{1}}{8:8}
] {pictures/data/mem-work/mi250-hip/throughput_vs_mem_mi250_hip_aligned_per_gpu.dat};

\addplot[name path=down8, draw=none, forget plot] table [
  x index=0,
  y expr=\thisrowno{2}-\thisrowno{3},
  restrict expr to domain={\thisrowno{1}}{8:8}
] {pictures/data/mem-work/mi250-hip/throughput_vs_mem_mi250_hip_aligned_per_gpu.dat};

\addplot[orange!90!black, fill opacity=0.18, forget plot] fill between[of=up8 and down8];

\addplot+[color=orange!90!black, solid, mark=diamond*] table [
  x index=0,
  y index=2,
  restrict expr to domain={\thisrowno{1}}{8:8}
] {pictures/data/mem-work/mi250-hip/throughput_vs_mem_mi250_hip_aligned_per_gpu.dat};
\addlegendentry{$W=8$}

\addplot[name path=up16, draw=none, forget plot] table [
  x index=0,
  y expr=\thisrowno{2}+\thisrowno{3},
  restrict expr to domain={\thisrowno{1}}{16:16}
] {pictures/data/mem-work/mi250-hip/throughput_vs_mem_mi250_hip_aligned_per_gpu.dat};

\addplot[name path=down16, draw=none, forget plot] table [
  x index=0,
  y expr=\thisrowno{2}-\thisrowno{3},
  restrict expr to domain={\thisrowno{1}}{16:16}
] {pictures/data/mem-work/mi250-hip/throughput_vs_mem_mi250_hip_aligned_per_gpu.dat};

\addplot[magenta!80!black, fill opacity=0.18, forget plot] fill between[of=up16 and down16];

\addplot+[color=magenta!80!black, solid, mark=pentagon*] table [
  x index=0,
  y index=2,
  restrict expr to domain={\thisrowno{1}}{16:16}
] {pictures/data/mem-work/mi250-hip/throughput_vs_mem_mi250_hip_aligned_per_gpu.dat};
\addlegendentry{$W=16$}

\nextgroupplot[
  ylabel={},
]

\addplot[name path=mbup1, draw=none] table [
  x index=0,
  y expr=\thisrowno{2}+\thisrowno{3},
  restrict expr to domain={\thisrowno{1}}{1:1}
] {pictures/data/mem-work/mi250-hip/throughput_vs_mem_mi250_hip_memory_only.dat};

\addplot[name path=mbdown1, draw=none] table [
  x index=0,
  y expr=\thisrowno{2}-\thisrowno{3},
  restrict expr to domain={\thisrowno{1}}{1:1}
] {pictures/data/mem-work/mi250-hip/throughput_vs_mem_mi250_hip_memory_only.dat};

\addplot[blue!80!black, fill opacity=0.18] fill between[of=mbup1 and mbdown1];

\addplot+[color=blue!80!black, solid, mark=*] table [
  x index=0,
  y index=2,
  restrict expr to domain={\thisrowno{1}}{1:1}
] {pictures/data/mem-work/mi250-hip/throughput_vs_mem_mi250_hip_memory_only.dat};

\addplot[name path=mbup2, draw=none] table [
  x index=0,
  y expr=\thisrowno{2}+\thisrowno{3},
  restrict expr to domain={\thisrowno{1}}{2:2}
] {pictures/data/mem-work/mi250-hip/throughput_vs_mem_mi250_hip_memory_only.dat};

\addplot[name path=mbdown2, draw=none] table [
  x index=0,
  y expr=\thisrowno{2}-\thisrowno{3},
  restrict expr to domain={\thisrowno{1}}{2:2}
] {pictures/data/mem-work/mi250-hip/throughput_vs_mem_mi250_hip_memory_only.dat};

\addplot[teal!80!black, fill opacity=0.18] fill between[of=mbup2 and mbdown2];

\addplot+[color=teal!80!black, solid, mark=square*] table [
  x index=0,
  y index=2,
  restrict expr to domain={\thisrowno{1}}{2:2}
] {pictures/data/mem-work/mi250-hip/throughput_vs_mem_mi250_hip_memory_only.dat};

\addplot[name path=mbup4, draw=none] table [
  x index=0,
  y expr=\thisrowno{2}+\thisrowno{3},
  restrict expr to domain={\thisrowno{1}}{4:4}
] {pictures/data/mem-work/mi250-hip/throughput_vs_mem_mi250_hip_memory_only.dat};

\addplot[name path=mbdown4, draw=none] table [
  x index=0,
  y expr=\thisrowno{2}-\thisrowno{3},
  restrict expr to domain={\thisrowno{1}}{4:4}
] {pictures/data/mem-work/mi250-hip/throughput_vs_mem_mi250_hip_memory_only.dat};

\addplot[green!50!black, fill opacity=0.18] fill between[of=mbup4 and mbdown4];

\addplot+[color=green!50!black, solid, mark=triangle*] table [
  x index=0,
  y index=2,
  restrict expr to domain={\thisrowno{1}}{4:4}
] {pictures/data/mem-work/mi250-hip/throughput_vs_mem_mi250_hip_memory_only.dat};

\addplot[name path=mbup8, draw=none] table [
  x index=0,
  y expr=\thisrowno{2}+\thisrowno{3},
  restrict expr to domain={\thisrowno{1}}{8:8}
] {pictures/data/mem-work/mi250-hip/throughput_vs_mem_mi250_hip_memory_only.dat};

\addplot[name path=mbdown8, draw=none] table [
  x index=0,
  y expr=\thisrowno{2}-\thisrowno{3},
  restrict expr to domain={\thisrowno{1}}{8:8}
] {pictures/data/mem-work/mi250-hip/throughput_vs_mem_mi250_hip_memory_only.dat};

\addplot[orange!90!black, fill opacity=0.18] fill between[of=mbup8 and mbdown8];

\addplot+[color=orange!90!black, solid, mark=diamond*] table [
  x index=0,
  y index=2,
  restrict expr to domain={\thisrowno{1}}{8:8}
] {pictures/data/mem-work/mi250-hip/throughput_vs_mem_mi250_hip_memory_only.dat};

\addplot[name path=mbup16, draw=none] table [
  x index=0,
  y expr=\thisrowno{2}+\thisrowno{3},
  restrict expr to domain={\thisrowno{1}}{16:16}
] {pictures/data/mem-work/mi250-hip/throughput_vs_mem_mi250_hip_memory_only.dat};

\addplot[name path=mbdown16, draw=none] table [
  x index=0,
  y expr=\thisrowno{2}-\thisrowno{3},
  restrict expr to domain={\thisrowno{1}}{16:16}
] {pictures/data/mem-work/mi250-hip/throughput_vs_mem_mi250_hip_memory_only.dat};

\addplot[magenta!80!black, fill opacity=0.18] fill between[of=mbup16 and mbdown16];

\addplot+[color=magenta!80!black, solid, mark=pentagon*] table [
  x index=0,
  y index=2,
  restrict expr to domain={\thisrowno{1}}{16:16}
] {pictures/data/mem-work/mi250-hip/throughput_vs_mem_mi250_hip_memory_only.dat};

\end{groupplot}
\end{tikzpicture}

%% file: pictures/tikz/mem_work_mi250_sycl.tex
\begin{tikzpicture}
\begin{groupplot}[
  group style={
    group size=2 by 1,
    horizontal sep=1cm,
    ylabels at=edge left,
    xlabels at=edge bottom
  },
  width=0.44\textwidth,
  height=0.44\textwidth,
  xmin=1, xmax=16,
  xmode=log,
  log basis x=2,
  xtick={1,2,4,8,16},
  xticklabels={1,2,4,8,16},
  xticklabel style={/pgf/number format/fixed},
  ymin=0, ymax=1500,
  ytick={0,500,1000,1500},
  xlabel={Memory Budget},
  x unit={\unit{\giga\byte}},
  ylabel={Throughput},
  y unit={ligands/\unit{\second}},
  grid=both,
  grid style={draw=gray!20},
  major grid style={draw=gray!30},
  unbounded coords=jump,
  tick label style={font=\footnotesize},
  label style={font=\footnotesize},
  title style={font=\footnotesize},
  legend style={font=\footnotesize, draw=none, fill=none, column sep=8pt},
  every axis plot/.append style={line width=1.1pt, mark size=2.2pt},
]

\nextgroupplot[
  legend to name=memworklegendmi250sycl,
  legend columns=3,
]

\addplot[name path=up1, draw=none, forget plot] table [
  x index=0,
  y expr=\thisrowno{2}+\thisrowno{3},
  restrict expr to domain={\thisrowno{1}}{1:1}
] {pictures/data/mem-work/mi250-sycl/throughput_vs_mem_mi250_sycl_aligned_per_gpu.dat};

\addplot[name path=down1, draw=none, forget plot] table [
  x index=0,
  y expr=\thisrowno{2}-\thisrowno{3},
  restrict expr to domain={\thisrowno{1}}{1:1}
] {pictures/data/mem-work/mi250-sycl/throughput_vs_mem_mi250_sycl_aligned_per_gpu.dat};

\addplot[blue!80!black, fill opacity=0.18, forget plot] fill between[of=up1 and down1];

\addplot+[color=blue!80!black, solid, mark=*] table [
  x index=0,
  y index=2,
  restrict expr to domain={\thisrowno{1}}{1:1}
] {pictures/data/mem-work/mi250-sycl/throughput_vs_mem_mi250_sycl_aligned_per_gpu.dat};
\addlegendentry{$W=1$}

\addplot[name path=up2, draw=none, forget plot] table [
  x index=0,
  y expr=\thisrowno{2}+\thisrowno{3},
  restrict expr to domain={\thisrowno{1}}{2:2}
] {pictures/data/mem-work/mi250-sycl/throughput_vs_mem_mi250_sycl_aligned_per_gpu.dat};

\addplot[name path=down2, draw=none, forget plot] table [
  x index=0,
  y expr=\thisrowno{2}-\thisrowno{3},
  restrict expr to domain={\thisrowno{1}}{2:2}
] {pictures/data/mem-work/mi250-sycl/throughput_vs_mem_mi250_sycl_aligned_per_gpu.dat};

\addplot[teal!80!black, fill opacity=0.18, forget plot] fill between[of=up2 and down2];

\addplot+[color=teal!80!black, solid, mark=square*] table [
  x index=0,
  y index=2,
  restrict expr to domain={\thisrowno{1}}{2:2}
] {pictures/data/mem-work/mi250-sycl/throughput_vs_mem_mi250_sycl_aligned_per_gpu.dat};
\addlegendentry{$W=2$}

\addplot[name path=up4, draw=none, forget plot] table [
  x index=0,
  y expr=\thisrowno{2}+\thisrowno{3},
  restrict expr to domain={\thisrowno{1}}{4:4}
] {pictures/data/mem-work/mi250-sycl/throughput_vs_mem_mi250_sycl_aligned_per_gpu.dat};

\addplot[name path=down4, draw=none, forget plot] table [
  x index=0,
  y expr=\thisrowno{2}-\thisrowno{3},
  restrict expr to domain={\thisrowno{1}}{4:4}
] {pictures/data/mem-work/mi250-sycl/throughput_vs_mem_mi250_sycl_aligned_per_gpu.dat};

\addplot[green!50!black, fill opacity=0.18, forget plot] fill between[of=up4 and down4];

\addplot+[color=green!50!black, solid, mark=triangle*] table [
  x index=0,
  y index=2,
  restrict expr to domain={\thisrowno{1}}{4:4}
] {pictures/data/mem-work/mi250-sycl/throughput_vs_mem_mi250_sycl_aligned_per_gpu.dat};
\addlegendentry{$W=4$}

\addplot[name path=up8, draw=none, forget plot] table [
  x index=0,
  y expr=\thisrowno{2}+\thisrowno{3},
  restrict expr to domain={\thisrowno{1}}{8:8}
] {pictures/data/mem-work/mi250-sycl/throughput_vs_mem_mi250_sycl_aligned_per_gpu.dat};

\addplot[name path=down8, draw=none, forget plot] table [
  x index=0,
  y expr=\thisrowno{2}-\thisrowno{3},
  restrict expr to domain={\thisrowno{1}}{8:8}
] {pictures/data/mem-work/mi250-sycl/throughput_vs_mem_mi250_sycl_aligned_per_gpu.dat};

\addplot[orange!90!black, fill opacity=0.18, forget plot] fill between[of=up8 and down8];

\addplot+[color=orange!90!black, solid, mark=diamond*] table [
  x index=0,
  y index=2,
  restrict expr to domain={\thisrowno{1}}{8:8}
] {pictures/data/mem-work/mi250-sycl/throughput_vs_mem_mi250_sycl_aligned_per_gpu.dat};
\addlegendentry{$W=8$}

\addplot[name path=up16, draw=none, forget plot] table [
  x index=0,
  y expr=\thisrowno{2}+\thisrowno{3},
  restrict expr to domain={\thisrowno{1}}{16:16}
] {pictures/data/mem-work/mi250-sycl/throughput_vs_mem_mi250_sycl_aligned_per_gpu.dat};

\addplot[name path=down16, draw=none, forget plot] table [
  x index=0,
  y expr=\thisrowno{2}-\thisrowno{3},
  restrict expr to domain={\thisrowno{1}}{16:16}
] {pictures/data/mem-work/mi250-sycl/throughput_vs_mem_mi250_sycl_aligned_per_gpu.dat};

\addplot[magenta!80!black, fill opacity=0.18, forget plot] fill between[of=up16 and down16];

\addplot+[color=magenta!80!black, solid, mark=pentagon*] table [
  x index=0,
  y index=2,
  restrict expr to domain={\thisrowno{1}}{16:16}
] {pictures/data/mem-work/mi250-sycl/throughput_vs_mem_mi250_sycl_aligned_per_gpu.dat};
\addlegendentry{$W=16$}

\nextgroupplot[
  ylabel={},
]

\addplot[name path=mbup1, draw=none] table [
  x index=0,
  y expr=\thisrowno{2}+\thisrowno{3},
  restrict expr to domain={\thisrowno{1}}{1:1}
] {pictures/data/mem-work/mi250-sycl/throughput_vs_mem_mi250_sycl_memory_only.dat};

\addplot[name path=mbdown1, draw=none] table [
  x index=0,
  y expr=\thisrowno{2}-\thisrowno{3},
  restrict expr to domain={\thisrowno{1}}{1:1}
] {pictures/data/mem-work/mi250-sycl/throughput_vs_mem_mi250_sycl_memory_only.dat};

\addplot[blue!80!black, fill opacity=0.18] fill between[of=mbup1 and mbdown1];

\addplot+[color=blue!80!black, solid, mark=*] table [
  x index=0,
  y index=2,
  restrict expr to domain={\thisrowno{1}}{1:1}
] {pictures/data/mem-work/mi250-sycl/throughput_vs_mem_mi250_sycl_memory_only.dat};

\addplot[name path=mbup2, draw=none] table [
  x index=0,
  y expr=\thisrowno{2}+\thisrowno{3},
  restrict expr to domain={\thisrowno{1}}{2:2}
] {pictures/data/mem-work/mi250-sycl/throughput_vs_mem_mi250_sycl_memory_only.dat};

\addplot[name path=mbdown2, draw=none] table [
  x index=0,
  y expr=\thisrowno{2}-\thisrowno{3},
  restrict expr to domain={\thisrowno{1}}{2:2}
] {pictures/data/mem-work/mi250-sycl/throughput_vs_mem_mi250_sycl_memory_only.dat};

\addplot[teal!80!black, fill opacity=0.18] fill between[of=mbup2 and mbdown2];

\addplot+[color=teal!80!black, solid, mark=square*] table [
  x index=0,
  y index=2,
  restrict expr to domain={\thisrowno{1}}{2:2}
] {pictures/data/mem-work/mi250-sycl/throughput_vs_mem_mi250_sycl_memory_only.dat};

\addplot[name path=mbup4, draw=none] table [
  x index=0,
  y expr=\thisrowno{2}+\thisrowno{3},
  restrict expr to domain={\thisrowno{1}}{4:4}
] {pictures/data/mem-work/mi250-sycl/throughput_vs_mem_mi250_sycl_memory_only.dat};

\addplot[name path=mbdown4, draw=none] table [
  x index=0,
  y expr=\thisrowno{2}-\thisrowno{3},
  restrict expr to domain={\thisrowno{1}}{4:4}
] {pictures/data/mem-work/mi250-sycl/throughput_vs_mem_mi250_sycl_memory_only.dat};

\addplot[green!50!black, fill opacity=0.18] fill between[of=mbup4 and mbdown4];

\addplot+[color=green!50!black, solid, mark=triangle*] table [
  x index=0,
  y index=2,
  restrict expr to domain={\thisrowno{1}}{4:4}
] {pictures/data/mem-work/mi250-sycl/throughput_vs_mem_mi250_sycl_memory_only.dat};

\addplot[name path=mbup8, draw=none] table [
  x index=0,
  y expr=\thisrowno{2}+\thisrowno{3},
  restrict expr to domain={\thisrowno{1}}{8:8}
] {pictures/data/mem-work/mi250-sycl/throughput_vs_mem_mi250_sycl_memory_only.dat};

\addplot[name path=mbdown8, draw=none] table [
  x index=0,
  y expr=\thisrowno{2}-\thisrowno{3},
  restrict expr to domain={\thisrowno{1}}{8:8}
] {pictures/data/mem-work/mi250-sycl/throughput_vs_mem_mi250_sycl_memory_only.dat};

\addplot[orange!90!black, fill opacity=0.18] fill between[of=mbup8 and mbdown8];

\addplot+[color=orange!90!black, solid, mark=diamond*] table [
  x index=0,
  y index=2,
  restrict expr to domain={\thisrowno{1}}{8:8}
] {pictures/data/mem-work/mi250-sycl/throughput_vs_mem_mi250_sycl_memory_only.dat};

\addplot[name path=mbup16, draw=none] table [
  x index=0,
  y expr=\thisrowno{2}+\thisrowno{3},
  restrict expr to domain={\thisrowno{1}}{16:16}
] {pictures/data/mem-work/mi250-sycl/throughput_vs_mem_mi250_sycl_memory_only.dat};

\addplot[name path=mbdown16, draw=none] table [
  x index=0,
  y expr=\thisrowno{2}-\thisrowno{3},
  restrict expr to domain={\thisrowno{1}}{16:16}
] {pictures/data/mem-work/mi250-sycl/throughput_vs_mem_mi250_sycl_memory_only.dat};

\addplot[magenta!80!black, fill opacity=0.18] fill between[of=mbup16 and mbdown16];

\addplot+[color=magenta!80!black, solid, mark=pentagon*] table [
  x index=0,
  y index=2,
  restrict expr to domain={\thisrowno{1}}{16:16}
] {pictures/data/mem-work/mi250-sycl/throughput_vs_mem_mi250_sycl_memory_only.dat};

\end{groupplot}
\end{tikzpicture}

%% file: pictures/tikz/mem_work_pvc_sycl.tex
\begin{tikzpicture}
\begin{groupplot}[
  group style={
    group size=2 by 1,
    horizontal sep=1cm,
    ylabels at=edge left,
    xlabels at=edge bottom
  },
  width=0.44\textwidth,
  height=0.44\textwidth,
  xmin=1, xmax=16,
  xmode=log,
  log basis x=2,
  xtick={1,2,4,8,16},
  xticklabels={1,2,4,8,16},
  xticklabel style={/pgf/number format/fixed},
  ymin=0, ymax=3800,
  ytick={0,1000,2000,3000},
  xlabel={Memory Budget},
  x unit={\unit{\giga\byte}},
  ylabel={Throughput},
  y unit={ligands/\unit{\second}},
  grid=both,
  grid style={draw=gray!20},
  major grid style={draw=gray!30},
  unbounded coords=jump,
  tick label style={font=\footnotesize},
  label style={font=\footnotesize},
  title style={font=\footnotesize},
  legend style={font=\footnotesize, draw=none, fill=none, column sep=8pt},
  every axis plot/.append style={line width=1.1pt, mark size=2.2pt},
]

\nextgroupplot[
  legend to name=memworklegendpvcsycl,
  legend columns=3,
]

\addplot[name path=up1, draw=none, forget plot] table [
  x index=0,
  y expr=\thisrowno{2}+\thisrowno{3},
  restrict expr to domain={\thisrowno{1}}{1:1}
] {pictures/data/mem-work/pvc-sycl/throughput_vs_mem_pvc_sycl_aligned_per_gpu.dat};

\addplot[name path=down1, draw=none, forget plot] table [
  x index=0,
  y expr=\thisrowno{2}-\thisrowno{3},
  restrict expr to domain={\thisrowno{1}}{1:1}
] {pictures/data/mem-work/pvc-sycl/throughput_vs_mem_pvc_sycl_aligned_per_gpu.dat};

\addplot[blue!80!black, fill opacity=0.18, forget plot] fill between[of=up1 and down1];

\addplot+[color=blue!80!black, solid, mark=*] table [
  x index=0,
  y index=2,
  restrict expr to domain={\thisrowno{1}}{1:1}
] {pictures/data/mem-work/pvc-sycl/throughput_vs_mem_pvc_sycl_aligned_per_gpu.dat};
\addlegendentry{$W=1$}

\addplot[name path=up2, draw=none, forget plot] table [
  x index=0,
  y expr=\thisrowno{2}+\thisrowno{3},
  restrict expr to domain={\thisrowno{1}}{2:2}
] {pictures/data/mem-work/pvc-sycl/throughput_vs_mem_pvc_sycl_aligned_per_gpu.dat};

\addplot[name path=down2, draw=none, forget plot] table [
  x index=0,
  y expr=\thisrowno{2}-\thisrowno{3},
  restrict expr to domain={\thisrowno{1}}{2:2}
] {pictures/data/mem-work/pvc-sycl/throughput_vs_mem_pvc_sycl_aligned_per_gpu.dat};

\addplot[teal!80!black, fill opacity=0.18, forget plot] fill between[of=up2 and down2];

\addplot+[color=teal!80!black, solid, mark=square*] table [
  x index=0,
  y index=2,
  restrict expr to domain={\thisrowno{1}}{2:2}
] {pictures/data/mem-work/pvc-sycl/throughput_vs_mem_pvc_sycl_aligned_per_gpu.dat};
\addlegendentry{$W=2$}

\addplot[name path=up4, draw=none, forget plot] table [
  x index=0,
  y expr=\thisrowno{2}+\thisrowno{3},
  restrict expr to domain={\thisrowno{1}}{4:4}
] {pictures/data/mem-work/pvc-sycl/throughput_vs_mem_pvc_sycl_aligned_per_gpu.dat};

\addplot[name path=down4, draw=none, forget plot] table [
  x index=0,
  y expr=\thisrowno{2}-\thisrowno{3},
  restrict expr to domain={\thisrowno{1}}{4:4}
] {pictures/data/mem-work/pvc-sycl/throughput_vs_mem_pvc_sycl_aligned_per_gpu.dat};

\addplot[green!50!black, fill opacity=0.18, forget plot] fill between[of=up4 and down4];

\addplot+[color=green!50!black, solid, mark=triangle*] table [
  x index=0,
  y index=2,
  restrict expr to domain={\thisrowno{1}}{4:4}
] {pictures/data/mem-work/pvc-sycl/throughput_vs_mem_pvc_sycl_aligned_per_gpu.dat};
\addlegendentry{$W=4$}

\addplot[name path=up8, draw=none, forget plot] table [
  x index=0,
  y expr=\thisrowno{2}+\thisrowno{3},
  restrict expr to domain={\thisrowno{1}}{8:8}
] {pictures/data/mem-work/pvc-sycl/throughput_vs_mem_pvc_sycl_aligned_per_gpu.dat};

\addplot[name path=down8, draw=none, forget plot] table [
  x index=0,
  y expr=\thisrowno{2}-\thisrowno{3},
  restrict expr to domain={\thisrowno{1}}{8:8}
] {pictures/data/mem-work/pvc-sycl/throughput_vs_mem_pvc_sycl_aligned_per_gpu.dat};

\addplot[orange!90!black, fill opacity=0.18, forget plot] fill between[of=up8 and down8];

\addplot+[color=orange!90!black, solid, mark=diamond*] table [
  x index=0,
  y index=2,
  restrict expr to domain={\thisrowno{1}}{8:8}
] {pictures/data/mem-work/pvc-sycl/throughput_vs_mem_pvc_sycl_aligned_per_gpu.dat};
\addlegendentry{$W=8$}

\addplot[name path=up16, draw=none, forget plot] table [
  x index=0,
  y expr=\thisrowno{2}+\thisrowno{3},
  restrict expr to domain={\thisrowno{1}}{16:16}
] {pictures/data/mem-work/pvc-sycl/throughput_vs_mem_pvc_sycl_aligned_per_gpu.dat};

\addplot[name path=down16, draw=none, forget plot] table [
  x index=0,
  y expr=\thisrowno{2}-\thisrowno{3},
  restrict expr to domain={\thisrowno{1}}{16:16}
] {pictures/data/mem-work/pvc-sycl/throughput_vs_mem_pvc_sycl_aligned_per_gpu.dat};

\addplot[magenta!80!black, fill opacity=0.18, forget plot] fill between[of=up16 and down16];

\addplot+[color=magenta!80!black, solid, mark=pentagon*] table [
  x index=0,
  y index=2,
  restrict expr to domain={\thisrowno{1}}{16:16}
] {pictures/data/mem-work/pvc-sycl/throughput_vs_mem_pvc_sycl_aligned_per_gpu.dat};
\addlegendentry{$W=16$}

\nextgroupplot[
  ylabel={},
]

\addplot[name path=mbup1, draw=none] table [
  x index=0,
  y expr=\thisrowno{2}+\thisrowno{3},
  restrict expr to domain={\thisrowno{1}}{1:1}
] {pictures/data/mem-work/pvc-sycl/throughput_vs_mem_pvc_sycl_memory_only.dat};

\addplot[name path=mbdown1, draw=none] table [
  x index=0,
  y expr=\thisrowno{2}-\thisrowno{3},
  restrict expr to domain={\thisrowno{1}}{1:1}
] {pictures/data/mem-work/pvc-sycl/throughput_vs_mem_pvc_sycl_memory_only.dat};

\addplot[blue!80!black, fill opacity=0.18] fill between[of=mbup1 and mbdown1];

\addplot+[color=blue!80!black, solid, mark=*] table [
  x index=0,
  y index=2,
  restrict expr to domain={\thisrowno{1}}{1:1}
] {pictures/data/mem-work/pvc-sycl/throughput_vs_mem_pvc_sycl_memory_only.dat};

\addplot[name path=mbup2, draw=none] table [
  x index=0,
  y expr=\thisrowno{2}+\thisrowno{3},
  restrict expr to domain={\thisrowno{1}}{2:2}
] {pictures/data/mem-work/pvc-sycl/throughput_vs_mem_pvc_sycl_memory_only.dat};

\addplot[name path=mbdown2, draw=none] table [
  x index=0,
  y expr=\thisrowno{2}-\thisrowno{3},
  restrict expr to domain={\thisrowno{1}}{2:2}
] {pictures/data/mem-work/pvc-sycl/throughput_vs_mem_pvc_sycl_memory_only.dat};

\addplot[teal!80!black, fill opacity=0.18] fill between[of=mbup2 and mbdown2];

\addplot+[color=teal!80!black, solid, mark=square*] table [
  x index=0,
  y index=2,
  restrict expr to domain={\thisrowno{1}}{2:2}
] {pictures/data/mem-work/pvc-sycl/throughput_vs_mem_pvc_sycl_memory_only.dat};

\addplot[name path=mbup4, draw=none] table [
  x index=0,
  y expr=\thisrowno{2}+\thisrowno{3},
  restrict expr to domain={\thisrowno{1}}{4:4}
] {pictures/data/mem-work/pvc-sycl/throughput_vs_mem_pvc_sycl_memory_only.dat};

\addplot[name path=mbdown4, draw=none] table [
  x index=0,
  y expr=\thisrowno{2}-\thisrowno{3},
  restrict expr to domain={\thisrowno{1}}{4:4}
] {pictures/data/mem-work/pvc-sycl/throughput_vs_mem_pvc_sycl_memory_only.dat};

\addplot[green!50!black, fill opacity=0.18] fill between[of=mbup4 and mbdown4];

\addplot+[color=green!50!black, solid, mark=triangle*] table [
  x index=0,
  y index=2,
  restrict expr to domain={\thisrowno{1}}{4:4}
] {pictures/data/mem-work/pvc-sycl/throughput_vs_mem_pvc_sycl_memory_only.dat};

\addplot[name path=mbup8, draw=none] table [
  x index=0,
  y expr=\thisrowno{2}+\thisrowno{3},
  restrict expr to domain={\thisrowno{1}}{8:8}
] {pictures/data/mem-work/pvc-sycl/throughput_vs_mem_pvc_sycl_memory_only.dat};

\addplot[name path=mbdown8, draw=none] table [
  x index=0,
  y expr=\thisrowno{2}-\thisrowno{3},
  restrict expr to domain={\thisrowno{1}}{8:8}
] {pictures/data/mem-work/pvc-sycl/throughput_vs_mem_pvc_sycl_memory_only.dat};

\addplot[orange!90!black, fill opacity=0.18] fill between[of=mbup8 and mbdown8];

\addplot+[color=orange!90!black, solid, mark=diamond*] table [
  x index=0,
  y index=2,
  restrict expr to domain={\thisrowno{1}}{8:8}
] {pictures/data/mem-work/pvc-sycl/throughput_vs_mem_pvc_sycl_memory_only.dat};

\addplot[name path=mbup16, draw=none] table [
  x index=0,
  y expr=\thisrowno{2}+\thisrowno{3},
  restrict expr to domain={\thisrowno{1}}{16:16}
] {pictures/data/mem-work/pvc-sycl/throughput_vs_mem_pvc_sycl_memory_only.dat};

\addplot[name path=mbdown16, draw=none] table [
  x index=0,
  y expr=\thisrowno{2}-\thisrowno{3},
  restrict expr to domain={\thisrowno{1}}{16:16}
] {pictures/data/mem-work/pvc-sycl/throughput_vs_mem_pvc_sycl_memory_only.dat};

\addplot[magenta!80!black, fill opacity=0.18] fill between[of=mbup16 and mbdown16];

\addplot+[color=magenta!80!black, solid, mark=pentagon*] table [
  x index=0,
  y index=2,
  restrict expr to domain={\thisrowno{1}}{16:16}
] {pictures/data/mem-work/pvc-sycl/throughput_vs_mem_pvc_sycl_memory_only.dat};

\end{groupplot}
\end{tikzpicture}

%% file: pictures/tikz/speedups_mixed.tex
\def\axisdefaultwidth{10cm}
\def\axisdefaultheight{4.5cm} 
\begin{tikzpicture}
\begin{axis}[
  width=0.96\textwidth,
  height=0.36\textwidth,
  ybar,
  bar width=10pt,
  ymin=0,
  ymax=7800,
  ytick={0,2000,4000,6000},
  xmin=-0.55,
  xmax=6.95,
  xtick={0,1,2,3.7,4.7,6.4},
  xticklabels={CUDA,HIP,SYCL,HIP,SYCL,SYCL},
  xticklabel style={font=\normalsize, align=center},
  extra x ticks={1,4.2,6.4},
  extra x tick labels={GH200,MI250,Max1550},
  extra x tick style={
    tick style={draw=none},
    major tick length=0pt,
    xticklabel style={font=\bfseries\normalsize, yshift=-15pt}
  },
  ymajorgrids=true,
  xmajorgrids=false,
  y grid style={draw=gray!30},
  tick label style={font=\normalsize},
  label style={font=\normalsize},
  ylabel={Throughput (lig/s)},
  enlarge x limits=false,
  clip=false,
  legend style={font=\normalsize, draw=none, fill=none, legend columns=2, at={(0.5,1.03)}, anchor=south},
]
\addplot+[
  fill=gray!45,
  draw=gray!70!black,
  bar shift=-5.5pt,
] table [col sep=space, x index=0, y index=1] {pictures/data/speedups/mixed-bars.dat};
\addlegendentry{Baseline}

\addplot+[
  fill=blue!55,
  draw=blue!80!black,
  bar shift=5.5pt,
  point meta=explicit,
  nodes near coords={\pgfmathprintnumber[fixed,precision=2]{\pgfplotspointmeta}x},
  every node near coord/.append style={font=\normalsize, anchor=south, yshift=0pt, text=black},
] table [col sep=space, x index=0, y index=2, meta index=3] {pictures/data/speedups/mixed-bars.dat};
\addlegendentry{Tuned}

\draw[gray!60, densely dotted, opacity=0.7, line width=1.15pt]
  ([yshift=-20pt]axis cs:2.85,0) -- (axis cs:2.85,7800);
\draw[gray!60, densely dotted, opacity=0.7, line width=1.15pt]
  ([yshift=-20pt]axis cs:5.55,0) -- (axis cs:5.55,7800);
\end{axis}
\end{tikzpicture}

%% file: pictures/tikz/pp_app.tex
\begin{tikzpicture}[x=1.05cm,y=0.78cm]
  \tikzstyle{cell}=[draw=white,line width=0.8pt,minimum width=1.08cm,minimum height=0.74cm,align=center,font=\footnotesize]
  \definecolor{effbad}{RGB}{201,76,64}
  \definecolor{effmid}{RGB}{236,208,122}
  \definecolor{effgood}{RGB}{79,160,104}

  \node[font=\bfseries\footnotesize] at (1, 3.35) {CUDA};
  \node[font=\bfseries\footnotesize] at (2, 3.35) {HIP};
  \node[font=\bfseries\footnotesize] at (3, 3.35) {SYCL};

  \node[anchor=east,font=\bfseries\footnotesize] at (0.20, 2.55) {GH200};
  \node[cell,fill={rgb,255:red,124;green,188;blue,116},text=black] at (1, 2.55) {0.99};
  \node[cell,fill={rgb,255:red,234;green,204;blue,130},text=black] at (3, 2.55) {0.52};

  \node[cell,fill={rgb,255:red,79;green,160;blue,104},text=white] at (2, 2.55) {1.00};
  \node[anchor=east,font=\bfseries\footnotesize] at (0.20, 1.65) {MI250};
  \node[cell,fill=gray!16,text=black] at (1, 1.65) {N/A};
  \node[cell,fill={rgb,255:red,142;green,194;blue,120},text=black] at (2, 1.65) {0.91};
  \node[cell,fill={rgb,255:red,79;green,160;blue,104},text=white] at (3, 1.65) {1.00};

  \node[anchor=east,font=\bfseries\footnotesize] at (0.20, 0.75) {Max1550};
  \node[cell,fill=gray!16,text=black] at (1, 0.75) {N/A};
  \node[cell,fill=gray!16,text=black] at (2, 0.75) {N/A};
  \node[cell,fill={rgb,255:red,79;green,160;blue,104},text=white] at (3, 0.75) {1.00};

  \draw[densely dashed,gray!60] (0.42, 0.30) -- (3.58, 0.30);

  \node[anchor=east,font=\bfseries\footnotesize] at (0.20, -0.25) {HM(eff)};
  \node[cell,fill={rgb,255:red,201;green,76;blue,64},text=white] at (1, -0.25) {0.00};
  \node[cell,fill={rgb,255:red,201;green,76;blue,64},text=white] at (2, -0.25) {0.00};
  \node[cell,fill={rgb,255:red,134;green,188;blue,114},text=black] at (3, -0.25) {0.76};

  \node[anchor=east,font=\bfseries\footnotesize] at (0.20, -1.10) {AM(eff)};
  \node[cell,fill={rgb,255:red,223;green,139;blue,92},text=white] at (1, -1.10) {0.33};
  \node[cell,fill={rgb,255:red,170;green,201;blue,124},text=black] at (2, -1.10) {0.64};
  \node[cell,fill={rgb,255:red,113;green,178;blue,108},text=black] at (3, -1.10) {0.84};

\end{tikzpicture}

%% file: sections/discussion.tex
\section{Results Discussion}
\label{sec:discussion}
In this section, we interpret the implications of the results, discuss the portability trade-offs revealed by the campaign, and explain the limitations of occupancy-derived heuristics as a basis for portable tuning.

\paragraph{Portability tradeoffs revealed by the campaign}
The cross-platform results also make clear that portability has more than one dimension.
The results measure practical end-to-end throughput, while the application-efficiency view normalizes each backend against the best result on the same platform.
Read together, these results show a recurring tradeoff: some backends achieve very strong local efficiency where they are available, but broader platform coverage still matters when the goal is portability across vendors.
This is why the paper reports both raw throughput and efficiency-based summaries rather than collapsing the discussion into a single number.

\paragraph{Limitations of occupancy-derived heuristics}
One consistent issue encountered in the data is that occupancy-derived quantities do not transfer cleanly across backends.
The occupancy-guided policy is reasonably well behaved on some platform/backend pairs, but its underlying heuristic is much less stable on others.
On Max1550 is the clearest example in this study, where the inferred occupancy multiple can become inflated relative to the values observed on GH200 and MI250.
For this reason, the paper should treat occupancy-guided batching as a useful implementation hint rather than as a portable first-principles rule.

\paragraph{Additional observations}
Beyond the main methodological result, the experiments also expose a secondary layer of runtime decisions that affect the throughput.
%
The bucket-multiplier results show that synchronization choices matter when multiple workers are competing for the same device. Removing kernel locking has little visible effect with one worker, but becomes beneficial as worker concurrency increases. 
This does not change the main conclusion of the paper, but it does show that finer-grained execution choices can widen or narrow the high-throughput operating region once the broader resource-management strategy is fixed.

\paragraph{Main takeaway from the results}
The clearest conclusion of this study is methodological rather than backend-specific.
Across the investigated vendor/framework combinations, the memory-bounded formulation consistently matches or exceeds the peak throughput of the occupancy-guided baseline.
These experimental results suggest that other irregular scientific applications can benefit from this type of partitioning to increase occupancy.
The same data also shows that portable throughput in repeated multi-kernel applications depends on out-of-kernel decisions such as work-unit formation, worker concurrency, and resource partitioning, not only on the performance of individual kernels viewed in isolation~\cite{accordi2024ook,yu2023unidock}.
On average, this methodology delivers a $1.5\times$ speedup across all tested platforms.

%% file: sections/related_work.tex
\section{Related Work}
\label{sec:related_work}

Molecular docking and virtual screening have long been driven by the tension between search accuracy and throughput.
Classical docking systems such as AutoDock4 and AutoDock Vina established the structure still used in many workflows: repeated evaluation of large ligand sets under irregular search dynamics~\cite{morris2009autodock4,trott2010vina,shoichet2004virtual,kitchen2004docking}.
Recent work has increasingly used HPC to speed up virtual screening campaigns and reduce their time to solution, while also targeting larger molecular libraries and more heterogeneous systems~\cite{legrand2020summit,gadioli2021exscalate,gadioli2024urgent_platform,lyu2019ultralarge,liu2025impact}.
This is the application context of our work. Our focus, however, is not docking accuracy or scoring-model design, but the systems problem of sustaining portable throughput across heterogeneous GPU backends.

Recent GPU docking work has made that systems dimension increasingly explicit.
AutoDock-GPU and related studies show that docking is not only compute intensive but also structurally irregular, with performance shaped by ligand-dependent search behavior, batching choices, and host-device interaction~\cite{santosmartins2021autodockgpu,solisvasquez2022irregular}.
Other modern GPU docking engines, such as Uni-Dock, push throughput further through aggressive batching and GPU-oriented dataflow design~\cite{yu2023unidock}, while comparative studies in earlier GPU docking work show that different mapping strategies can shift the balance between latency and throughput~\cite{vitali2024gpu_optimized_docking}.
Our paper is closest to this systems-oriented branch of docking research, but differs in focus: rather than proposing a new fastest docking engine, we study how batching and worker-level resource decisions transfer across CUDA, HIP, and SYCL implementations.

This work builds on the out-of-kernel perspective introduced in prior docking studies~\cite{accordi2024ook,vitali2024gpu_optimized_docking}. 
The main idea is that performance portability is shaped by policy choices outside the kernels themselves, such as bucketing, batching, and stream-level concurrency.
These choices are especially delicate on heterogeneous GPU platforms, where occupancy information, register usage, and runtime behavior differ across CUDA, HIP, and SYCL implementations~\cite{pennycook2019implications,pennycook2021revisiting,crisci2024ligen_pipeline_pp}.
The most direct methodological precedent is prior out-of-kernel docking work, which showed that docking throughput can depend strongly on bucket sizing and other host-side decisions outside the main kernels~\cite{accordi2024ook}.
Follow-on portability studies of that pipeline have shown that these host-side decisions are entangled with backend-dependent effects such as occupancy estimation, compiler behavior, and register pressure~\cite{crisci2024ligen_pipeline_pp,accordi2024iwocl_lumig}.
This literature is important for our paper because it shows that performance depends not only on kernel code, but also on how work is formed and executed across the full pipeline.
Our work builds on that point, but treats these choices as part of a more general optimization methodology for repeated multi-kernel GPU pipelines rather than as a docking-specific optimization.

This broader framing also connects to the mini-app and proxy-app tradition in HPC, where compact applications are used to expose performance-critical behavior without the full complexity of a production workflow~\cite{heroux2009improving,thavappiragasam2020minimdock,poenaru2021ppmodels}.
From that perspective, \texttt{muDock} plays the role of a representative repeated multi-kernel scientific application: it preserves the irregularity and orchestration structure that matter to optimization, while remaining small enough to study systematically across platforms and backends.
Previous work has already used \texttt{muDock} to study vectorization-oriented performance portability on CPUs~\cite{accordi2025mudock_cpu}; the present paper extends that line of work toward GPU-oriented out-of-kernel optimization and backend portability.

Finally, our study is embedded in the wider performance-portability literature.
Pennycook and coauthors formalized performance portability as a joint requirement of efficiency and portability, and follow-on work has emphasized that evaluating portability requires more than demonstrating functional correctness across programming models~\cite{pennycook2019implications,pennycook2021revisiting}.
Subsequent cross-platform studies have reinforced this point for both full applications and benchmarking suites, including recent work on programming-model comparisons and on SYCL behavior across NVIDIA, AMD, and Intel platforms~\cite{deakin2021yellowbrick,reguly2023sycl,jin2023syclbench}.
Our contribution is deliberately narrower and more application-driven: we isolate one consequential out-of-kernel policy question in a repeated multi-kernel scientific workflow and show how a portable methodology can improve throughput while also revealing the remaining limits imposed by backend-specific behavior.

%% file: sections/conclusion.tex
\section{Conclusion}
\label{sec:conclusion}

This paper presented a portable out-of-kernel optimization methodology for scientific applications organized as repeated multi-kernel GPU pipelines, and studied it through the \texttt{muDock} molecular docking mini-app.
More generally, the paper shows that for similar irregular throughput-oriented workloads, application throughput can be improved by tuning three coupled runtime dimensions: the memory allocated to each worker, the batching policy used to form the next execution chunk, and the number of workers that concurrently feed the device.

Across the investigated platforms and backends, this methodology improves the mixed-workload throughput by 1.53$\times$ on average over the monolithic reference configuration.
At the policy level, the memory-bounded batching method matches or exceeds the peak throughput of the occupancy-guided method on every evaluated platform/backend pair.
These results indicate that tuning memory allocation, batching, and worker concurrency together provides a practical way to improve throughput while preserving portability across CUDA, HIP, and SYCL implementations.

The broader lesson is that repeated multi-kernel pipelines should be tuned at application level, not only at kernel level.
For example, for the use case under consideration, the proposed methodology delivers a significant speedup, meaning that large-scale virtual screening setups can analyze a larger chemical space in the same amount of time budget~\cite{lyu2019ultralarge}.
In this setting, explicit resource partitioning and concurrent work submission provide a stronger basis for portable performance than occupancy-derived alignment alone, even though backend-dependent behavior still limits full portability.


%% file: sections/appendix_plots.tex
\section{Supplementary Plots}
\label{sec:appendix_plots}

This appendix collects the experiment plots omitted from the main text for space reasons.
The representative figures discussed in Section~\ref{sec:experimental_result} remain the primary evidence for the paper narrative; the plots below provide the complete comparison set for the remaining platform/backend pairs.

\subsection{Complete Worker-Memory Sweeps}

\begin{figure*}[t]
\centering
\includestandalone[width=\linewidth]{pictures/tikz/mem_work_gh200_hip}
\caption{Worker-memory sensitivity on GH200/HIP. Memory-bounded batching reaches the highest throughput over the broadest region of the sweep.}
\label{fig:app-gh200-hip}
\end{figure*}

\begin{figure*}[t]
\centering
\includestandalone[width=\linewidth]{pictures/tikz/mem_work_gh200_sycl}
\caption{Worker-memory sensitivity on GH200/SYCL. Memory-bounded batching preserves the better plateau across worker and memory settings.}
\label{fig:app-gh200-sycl}
\end{figure*}

\begin{figure*}[t]
\centering
\includestandalone[width=\linewidth]{pictures/tikz/mem_work_mi250_sycl}
\caption{Worker-memory sensitivity on MI250/SYCL. Memory-bounded batching broadens the high-throughput region and reaches the best peak.}
\label{fig:app-mi250-sycl}
\end{figure*}

\subsection{Complete Bucket-Multiplier Sweeps}

\begin{figure*}[t]
\centering
\includestandalone[width=\linewidth]{pictures/tikz/bucket_mult_gh200_hip}
\caption{Bucket-multiplier sensitivity on GH200/HIP. Concurrent GPU access broadens the best-performing region once multiple workers feed the same device.}
\label{fig:app-bucket-mult-gh200-hip}
\end{figure*}

\begin{figure*}[t]
\centering
\includestandalone[width=\linewidth]{pictures/tikz/bucket_mult_gh200_sycl}

\caption{Bucket-multiplier sensitivity on GH200/SYCL. Concurrent GPU access remains most relevant once the higher-worker regimes become active.}
\label{fig:app-bucket-mult-gh200-sycl}
\end{figure*}

\begin{figure*}[t]
\centering
\includestandalone[width=\linewidth]{pictures/tikz/bucket_mult_mi250_hip}
\caption{Bucket-multiplier sensitivity on MI250/HIP. Concurrent GPU access provides the larger high-throughput window as concurrency increases.}
\label{fig:app-bucket-mult-mi250-hip}
\end{figure*}

\begin{figure*}[t]
\centering
\includestandalone[width=\linewidth]{pictures/tikz/bucket_mult_mi250_sycl}
\caption{Bucket-multiplier sensitivity on MI250/SYCL. Concurrent GPU access modestly improves the stable operating region under higher worker counts.}
\label{fig:app-bucket-mult-mi250-sycl}
\end{figure*}

\begin{figure*}[t]
\centering
\includestandalone[width=\linewidth]{pictures/tikz/bucket_mult_pvc_sycl}
\caption{Bucket-multiplier sensitivity on PVC/SYCL. The two access modes remain closer than on GH200, but concurrent GPU access still offers the cleaner plateau.}
\label{fig:app-bucket-mult-pvc-sycl}
\end{figure*}

\subsection{Supplementary Register Summaries}

The latest review bundle also includes refreshed register-summary tables for the GH200/CUDA, GH200/HIP, MI250/HIP, MI250/SYCL, and PVC/SYCL backends.
These summaries are used in the main text only as supporting evidence for the backend-specific interpretation.
They are intentionally not treated as a complete root-cause analysis because the current parser reports coarse generic metrics and still lacks GH200/SYCL support.

%% file: pictures/tikz/bucket_mult_pvc_sycl.tex
\begin{tikzpicture}
\begin{groupplot}[
  group style={
    group size=2 by 1,
    horizontal sep=1cm,
    ylabels at=edge left,
    xlabels at=edge bottom
  },
  width=0.44\textwidth,
  height=0.44\textwidth,
  xmin=1, xmax=7,
  xtick={1,2,3,4,5,6,7},
  xticklabels={1,2,3,4,5,6,7},
  xticklabel style={/pgf/number format/fixed},
  ymin=0, ymax=4000,
  ytick={0,1000,2000,3000, 4000},
  xlabel={Memory Budget},
  x unit={\unit{\giga\byte}},
  ylabel={Throughput},
  y unit={ligands/\unit{\second}},
  grid=both,
  grid style={draw=gray!20},
  major grid style={draw=gray!30},
  unbounded coords=jump,
  tick label style={font=\footnotesize},
  label style={font=\footnotesize},
  title style={font=\footnotesize},
  legend style={font=\footnotesize, draw=none, fill=none, column sep=8pt},
  every axis plot/.append style={line width=1.1pt, mark size=2.2pt},
]

\nextgroupplot[
  legend to name=memworklegendpvcsycl,
  legend columns=3,
]

\addplot[name path=up1, draw=none, forget plot] table [
  x index=0,
  y expr=\thisrowno{2}+\thisrowno{3},
  restrict expr to domain={\thisrowno{1}}{1:1}
] {pictures/data/bucket-mult/pvc-sycl/throughput_vs_bucket_pvc_sycl_kernel_lock_off.dat};

\addplot[name path=down1, draw=none, forget plot] table [
  x index=0,
  y expr=\thisrowno{2}-\thisrowno{3},
  restrict expr to domain={\thisrowno{1}}{1:1}
] {pictures/data/bucket-mult/pvc-sycl/throughput_vs_bucket_pvc_sycl_kernel_lock_off.dat};

\addplot[blue!80!black, fill opacity=0.18, forget plot] fill between[of=up1 and down1];

\addplot+[color=blue!80!black, solid, mark=*] table [
  x index=0,
  y index=2,
  restrict expr to domain={\thisrowno{1}}{1:1}
] {pictures/data/bucket-mult/pvc-sycl/throughput_vs_bucket_pvc_sycl_kernel_lock_off.dat};
\addlegendentry{$W=1$}

\addplot[name path=up2, draw=none, forget plot] table [
  x index=0,
  y expr=\thisrowno{2}+\thisrowno{3},
  restrict expr to domain={\thisrowno{1}}{2:2}
] {pictures/data/bucket-mult/pvc-sycl/throughput_vs_bucket_pvc_sycl_kernel_lock_off.dat};

\addplot[name path=down2, draw=none, forget plot] table [
  x index=0,
  y expr=\thisrowno{2}-\thisrowno{3},
  restrict expr to domain={\thisrowno{1}}{2:2}
] {pictures/data/bucket-mult/pvc-sycl/throughput_vs_bucket_pvc_sycl_kernel_lock_off.dat};

\addplot[teal!80!black, fill opacity=0.18, forget plot] fill between[of=up2 and down2];

\addplot+[color=teal!80!black, solid, mark=square*] table [
  x index=0,
  y index=2,
  restrict expr to domain={\thisrowno{1}}{2:2}
] {pictures/data/bucket-mult/pvc-sycl/throughput_vs_bucket_pvc_sycl_kernel_lock_off.dat};
\addlegendentry{$W=2$}

\addplot[name path=up4, draw=none, forget plot] table [
  x index=0,
  y expr=\thisrowno{2}+\thisrowno{3},
  restrict expr to domain={\thisrowno{1}}{4:4}
] {pictures/data/bucket-mult/pvc-sycl/throughput_vs_bucket_pvc_sycl_kernel_lock_off.dat};

\addplot[name path=down4, draw=none, forget plot] table [
  x index=0,
  y expr=\thisrowno{2}-\thisrowno{3},
  restrict expr to domain={\thisrowno{1}}{4:4}
] {pictures/data/bucket-mult/pvc-sycl/throughput_vs_bucket_pvc_sycl_kernel_lock_off.dat};

\addplot[green!50!black, fill opacity=0.18, forget plot] fill between[of=up4 and down4];

\addplot+[color=green!50!black, solid, mark=triangle*] table [
  x index=0,
  y index=2,
  restrict expr to domain={\thisrowno{1}}{4:4}
] {pictures/data/bucket-mult/pvc-sycl/throughput_vs_bucket_pvc_sycl_kernel_lock_off.dat};
\addlegendentry{$W=4$}

\addplot[name path=up8, draw=none, forget plot] table [
  x index=0,
  y expr=\thisrowno{2}+\thisrowno{3},
  restrict expr to domain={\thisrowno{1}}{8:8}
] {pictures/data/bucket-mult/pvc-sycl/throughput_vs_bucket_pvc_sycl_kernel_lock_off.dat};

\addplot[name path=down8, draw=none, forget plot] table [
  x index=0,
  y expr=\thisrowno{2}-\thisrowno{3},
  restrict expr to domain={\thisrowno{1}}{8:8}
] {pictures/data/bucket-mult/pvc-sycl/throughput_vs_bucket_pvc_sycl_kernel_lock_off.dat};

\addplot[orange!90!black, fill opacity=0.18, forget plot] fill between[of=up8 and down8];

\addplot+[color=orange!90!black, solid, mark=diamond*] table [
  x index=0,
  y index=2,
  restrict expr to domain={\thisrowno{1}}{8:8}
] {pictures/data/bucket-mult/pvc-sycl/throughput_vs_bucket_pvc_sycl_kernel_lock_off.dat};
\addlegendentry{$W=8$}

\nextgroupplot[
  ylabel={},
]

\addplot[name path=mbup1, draw=none] table [
  x index=0,
  y expr=\thisrowno{2}+\thisrowno{3},
  restrict expr to domain={\thisrowno{1}}{1:1}
] {pictures/data/bucket-mult/pvc-sycl/throughput_vs_bucket_pvc_sycl_kernel_lock_on.dat};

\addplot[name path=mbdown1, draw=none] table [
  x index=0,
  y expr=\thisrowno{2}-\thisrowno{3},
  restrict expr to domain={\thisrowno{1}}{1:1}
] {pictures/data/bucket-mult/pvc-sycl/throughput_vs_bucket_pvc_sycl_kernel_lock_on.dat};

\addplot[blue!80!black, fill opacity=0.18] fill between[of=mbup1 and mbdown1];

\addplot+[color=blue!80!black, solid, mark=*] table [
  x index=0,
  y index=2,
  restrict expr to domain={\thisrowno{1}}{1:1}
] {pictures/data/bucket-mult/pvc-sycl/throughput_vs_bucket_pvc_sycl_kernel_lock_on.dat};

\addplot[name path=mbup2, draw=none] table [
  x index=0,
  y expr=\thisrowno{2}+\thisrowno{3},
  restrict expr to domain={\thisrowno{1}}{2:2}
] {pictures/data/bucket-mult/pvc-sycl/throughput_vs_bucket_pvc_sycl_kernel_lock_on.dat};

\addplot[name path=mbdown2, draw=none] table [
  x index=0,
  y expr=\thisrowno{2}-\thisrowno{3},
  restrict expr to domain={\thisrowno{1}}{2:2}
] {pictures/data/bucket-mult/pvc-sycl/throughput_vs_bucket_pvc_sycl_kernel_lock_on.dat};

\addplot[teal!80!black, fill opacity=0.18] fill between[of=mbup2 and mbdown2];

\addplot+[color=teal!80!black, solid, mark=square*] table [
  x index=0,
  y index=2,
  restrict expr to domain={\thisrowno{1}}{2:2}
] {pictures/data/bucket-mult/pvc-sycl/throughput_vs_bucket_pvc_sycl_kernel_lock_on.dat};

\addplot[name path=mbup4, draw=none] table [
  x index=0,
  y expr=\thisrowno{2}+\thisrowno{3},
  restrict expr to domain={\thisrowno{1}}{4:4}
] {pictures/data/bucket-mult/pvc-sycl/throughput_vs_bucket_pvc_sycl_kernel_lock_on.dat};

\addplot[name path=mbdown4, draw=none] table [
  x index=0,
  y expr=\thisrowno{2}-\thisrowno{3},
  restrict expr to domain={\thisrowno{1}}{4:4}
] {pictures/data/bucket-mult/pvc-sycl/throughput_vs_bucket_pvc_sycl_kernel_lock_on.dat};

\addplot[green!50!black, fill opacity=0.18] fill between[of=mbup4 and mbdown4];

\addplot+[color=green!50!black, solid, mark=triangle*] table [
  x index=0,
  y index=2,
  restrict expr to domain={\thisrowno{1}}{4:4}
] {pictures/data/bucket-mult/pvc-sycl/throughput_vs_bucket_pvc_sycl_kernel_lock_on.dat};

\addplot[name path=mbup8, draw=none] table [
  x index=0,
  y expr=\thisrowno{2}+\thisrowno{3},
  restrict expr to domain={\thisrowno{1}}{8:8}
] {pictures/data/bucket-mult/pvc-sycl/throughput_vs_bucket_pvc_sycl_kernel_lock_on.dat};

\addplot[name path=mbdown8, draw=none] table [
  x index=0,
  y expr=\thisrowno{2}-\thisrowno{3},
  restrict expr to domain={\thisrowno{1}}{8:8}
] {pictures/data/bucket-mult/pvc-sycl/throughput_vs_bucket_pvc_sycl_kernel_lock_on.dat};

\addplot[orange!90!black, fill opacity=0.18] fill between[of=mbup8 and mbdown8];

\addplot+[color=orange!90!black, solid, mark=diamond*] table [
  x index=0,
  y index=2,
  restrict expr to domain={\thisrowno{1}}{8:8}
] {pictures/data/bucket-mult/pvc-sycl/throughput_vs_bucket_pvc_sycl_kernel_lock_on.dat};

\end{groupplot}
\end{tikzpicture}